\documentclass[aps,prl,showpacs,amsmath,amssymb,amsfonts,superscriptaddress,tightenlines,twocolumn,lengthcheck]{revtex4-1}
\usepackage{amsmath}
\usepackage{amsfonts}
\usepackage{graphicx}
\usepackage{epsfig}
\usepackage{color}
\usepackage[colorlinks,citecolor=blue]{hyperref}

\begin{document}

\title{Generalized Ultrastrong Optomechanics}
\author{Jie-Qiao Liao}
\email{jqliao@hunnu.edu.cn}
\affiliation{Key Laboratory of Low-Dimensional Quantum Structures and Quantum Control of
Ministry of Education, Department of Physics and Synergetic Innovation
Center for Quantum Effects and Applications, Hunan Normal University,
Changsha 410081, China}
\author{Jin-Feng Huang}
\affiliation{Key Laboratory of Low-Dimensional Quantum Structures and Quantum Control of
Ministry of Education, Department of Physics and Synergetic Innovation
Center for Quantum Effects and Applications, Hunan Normal University,
Changsha 410081, China}
\author{Lin Tian}
\affiliation{School of Natural Sciences, University of California, Merced, California 95343, USA}
\author{Le-Man Kuang}
\affiliation{Key Laboratory of Low-Dimensional Quantum Structures and Quantum Control of
Ministry of Education, Department of Physics and Synergetic Innovation
Center for Quantum Effects and Applications, Hunan Normal University,
Changsha 410081, China}
\author{Chang-Pu Sun}
\affiliation{Beijing Computational Science Research Center, Beijing 100193, China}
\affiliation{Graduate School of Chinese Academy of Engineering Physics, Beijing 100084, China}

\begin{abstract}
We propose a reliable scheme to realize a generalized ultrastrong optomechanical coupling in a two-mode cross-Kerr-type coupled system, where one of the bosonic modes is strongly driven. The effective optomechanical interaction takes the form of a product of the photon number operator of one mode and the quadrature operator of the other mode. The coupling strength and quadrature phase are both tunable via the driving field. The coupling strength can be strongly enhanced to reach the ultrastrong-coupling regime, where the few-photon optomechanical effects such as photon blockade and macroscopically distinct quantum superposition become accessible. The presence of tunable quadrature phase also enables the implementation of geometric quantum operations. Numerical simulations show that this method works well in a wide parameter space. We also present an analysis of the experimental implementation of this scheme.
\end{abstract}

\date{\today}
\maketitle

\emph{Introduction.}---Light-matter interaction is at the heart of cavity optomechanics~\cite{Kippenberg2008rev,Aspelmeyer2012rev,Aspelmeyer2014} and is the root of various quantum coherence effects in optomechanical systems. The studies of cavity optomechanics focus primarily on the understanding, manipulation, and exploitation of the optomechanical couplings, and aim to explore both the fundamentals of quantum theory and modern quantum technology. Of particular interest is the study of optomechanics at the few-photon level~\cite{Nunnenkamp2011,Rabl2011,Liao2012,Liao2013,Hong2013,Xu2013,Marshall2003,Liao2016}. This is because the nonlinear optomechanical interaction is an intrinsic characteristic of optomechanics. Many interesting effects appear in this regime, such as phonon sideband spectrum~\cite{Nunnenkamp2011,Liao2012}, photon blockade in the cavity driven by a continuous wave~\cite{Rabl2011} or a wave packet~\cite{Liao2013}, and macroscopic quantum coherence~\cite{Marshall2003,Liao2016}. However, the few-photon optomechanical effects have not been observed in experiments because the single-photon optomechanical coupling is too weak to be resolved from the environmental noise. How to enhance the optomechanical coupling remains an important challenge in this field. Until now, people proposed several methods to enhance the single-photon optomechanical coupling. These methods include the construction of an array of mechanical resonators~\cite{Xuereb2012}, the use of the nonlinearity in Josephson junctions~\cite{Rimberg2014,Heikkila2014,Pirkkalainen2015}, the modulation of the couplings~\cite{Liao2014}, and the utilization of quantum squeezing resources~\cite{Lue2015,Li2016}, and mechanical amplification~\cite{Lemonde2016}.

In this Letter, we propose an efficient approach to realize ultrastrong optomechanical coupling in the few-photon regime~\cite{seeSM}. Here ultrastrong coupling is defined as the strength of the single-photon optomechanical coupling is a considerable fraction of the mechanical frequency~\cite{Hu2015}. Our scheme is realized by applying strong driving on one of the two bosonic modes coupled by the cross-Kerr interaction. Note that the cross-Kerr interaction has been widely used in quantum state preparation~\cite{Paternostro2003,Kuang2003}, quantum information protocols~\cite{Milburn1989PRL,Vitali2000,Kok2007,Nemoto2004}, quantum nondemolition photon measurement~\cite{Imoto1985,Grangier1998}, and phonon counting~\cite{Ding2017}. In particular, the generalized optomechanical coupling takes the form of the product of the occupation number operator of one mode and the quadrature operator of the other mode. Here, the strength of single-photon optomechanical coupling is enhanced by the driving to reach the single-photon strong coupling regime. Our scheme has the following features. (i) The driving field enhances the optomechanical coupling strength to reach the ultrastrong coupling regime, and the generalized optomechanical coupling can be used to implement geometric quantum operations with proper quadrature angle sequences. (ii) This method works for both steady-state and transient displacements, which correspond to constant and modulated optomechanical coupling cases. (iii) In the displacement representation, the driving detuning plays the role of the effective mechanical frequency, and hence it is possible to choose a high natural frequency of the mechanical mode to suppress its thermal noise.

\emph{Model.}---We consider two bosonic modes $a$ and $b$ coupled by a cross-Kerr interaction. One of the modes (for instance mode $b$) is driven by a monochromatic field with frequency $\omega_{Lb}$. In a rotating frame with respect to $H_{0}=\omega_{Lb}b^{\dagger}b$, the Hamiltonian of this system reads ($\hbar=1$)
\begin{equation}
H_{I}=\omega_{a}a^{\dagger}a+\Delta_{b}b^{\dagger}b+\chi a^{\dagger}ab^{\dagger}b+\Omega_{b}b^{\dagger}+\Omega_{b}^{\ast}b,
\end{equation}
where $a$ ($a^{\dagger}$) and $b$ ($b^{\dagger}$) are the annihilation (creation) operators of the two bosonic modes,
with the corresponding resonance frequencies $\omega_{a}$ and $\omega_{b}$. The parameter $\Delta_{b}=\omega_{b}-\omega_{Lb}$ is the detuning of the resonance frequency of mode $b$ with respect to the driving frequency $\omega_{Lb}$, and the parameter $\Omega_{b}$ is the driving amplitude. The two modes are coupled to each other through a cross-Kerr interaction, with the coupling strength $\chi$.

To treat the damping and noise in this system, we assume that the two bosonic modes are coupled to two independent Markovian
environments, the evolution of the system is hence governed by the quantum master equation
\begin{eqnarray}
\dot{\rho}&=&i[\rho, H_{I}]+\gamma_{a}(\bar{n}_{a}+1)\mathcal{D}[a]\rho+\gamma_{a}\bar{n}_{a}\mathcal{D}[a^{\dag}]\rho\nonumber\\
&&+\gamma_{b}(\bar{n}_{b}+1)\mathcal{D}[b]\rho+\gamma_{b}\bar{n}_{b}\mathcal{D}[b^{\dag}]\rho,\label{mastereqoriMT}
\end{eqnarray}
where $\mathcal{D}[o]\rho=o\rho o^{\dagger}-(o^{\dagger}o\rho+\rho o^{\dagger}o)/2$ is the standard Lindblad superoperator for bosonic-mode damping, $\gamma_{a}$ ($\gamma_{b}$) and $\bar{n}_{a}$ ($\bar{n}_{b}$) are the damping rate and environment thermal excitation occupation of mode $a$ ($b$), respectively.

\emph{Generalized ultrastrong coupling.}---Our motivation in this work is to obtain an ultrastrong optomechanical coupling between the two modes. Under strong driving, the mode $b$ is excited with large occupation number, and the operator $b$ can be written as a summation of its mean value and a quantum fluctuation $b\rightarrow\beta+b$, and similarly $b^{\dagger}\rightarrow\beta^{\ast}+b^{\dagger}$. Note that the occupation number of mode $a$ is independent of the driving on mode $b$ because the operator $a^{\dag}a$ is a conserved quantity. The cross-Kerr interaction then becomes $\chi a^{\dagger}a(\beta^{\ast}+b^{\dagger})(\beta+b)=\chi\beta^{\ast}\beta a^{\dagger}a+\chi a^{\dagger}a(\beta^{\ast}b+\beta b^{\dagger})+\chi a^{\dagger}ab^{\dagger}b$. Here the first term is a frequency shift on mode $a$, the second term is the generalized optomechanical coupling with a coupling strength enhanced by a factor $|\beta|$, and the third term is the cross-Kerr interaction between mode $a$ and the fluctuation of mode $b$.

To prove the above analysis, we perform the transformation $\rho'=D_{b}(\beta)\rho D_{b}^{\dagger}(\beta)$ to the quantum master equation~(\ref{mastereqoriMT}), where $\beta=\vert\beta\vert e^{i\theta}$ is the mean displacement of mode $b$. By performing this transformation, we  obtain the equation of motion of the displacement as
$\dot{\beta}=-(i\Delta_{b}+\gamma_{b}/2)\beta+i\Omega_{b}$. We consider the case where the time scale of system relaxation is much shorter than other time scales. The steady-state displacement reads $\beta_{\textrm{ss}}=\Omega_{b}/(\Delta_{b}-i\gamma_{b}/2)$, which is a tunable complex number by choosing proper $\Omega_{b}$ and $\Delta_{b}$. In the displacement representation, the quantum master equation takes the same form as Eq.~(\ref{mastereqoriMT}) under the replacement $\rho\rightarrow\rho'$ and $H_{I}\rightarrow H_{\textrm{tra}}$, where the transformed Hamiltonian is given by
$H_{\textrm{tra}}=\omega_{a}^{\prime}a^{\dagger}a+\Delta_{b}b^{\dagger}b-\chi a^{\dagger}a(\beta_{\textrm{ss}}b^{\dagger}+\beta_{\textrm{ss}}^{\ast}b)+\chi a^{\dagger}ab^{\dagger}b$, with the frequency $\omega_{a}^{\prime}=\omega_{a}+\chi|\beta_{\textrm{ss}}|^{2}$.
In this Hamiltonian, the cross-Kerr term is an effective frequency shift for the two modes. When this frequency shift $m\chi$ associated with $m$ excitations in mode $a$ is much smaller than the effective frequency of mode $b$, namely $m\chi\ll\Delta_{b}$, we can neglect the cross-Kerr interaction term safely. In this case, a generalized ultrastrong optomechanical interaction can be obtained.

In this work, we focus on few-photon optomechanics and hence consider the regime $\chi\ll\Delta_{b}$ and $\chi\vert\beta_{\textrm{ss}}\vert\sim \Delta_{b}$, then the cross-Kerr interaction term in $H_{\textrm{tra}}$ can be safely discarded and we obtain the generalized optomechanical Hamiltonian
\begin{eqnarray}
H_{\textrm{app}}=\omega_{a}^{\prime}a^{\dagger}a+\Delta_{b}b^{\dagger}b-g_{0}a^{\dagger}a(b^{\dagger}e^{i\theta}+be^{-i\theta}),\label{HapproxedMT}
\end{eqnarray}
where $g_{0}=\chi\vert\beta_{\textrm{ss}}\vert$ is the single-photon optomechanical coupling strength. The Hamiltonian~(\ref{HapproxedMT}) possesses three features: (i) The effective resonance frequency $\Delta_{b}$ of the mechanical mode is tunable by choosing proper driving frequency $\omega_{Lb}$. Therefore, we can choose a small $\Delta_{b}$ such that a near-resonant displacement interaction is obtained and further the displacement effect of single photons is enhanced. (ii) The single-photon optomechanical coupling between the two modes is enhanced
by a factor of the displacement amplitude $|\beta_{\textrm{ss}}|$, which is determined by the driving amplitude $\Omega_{b}$. Therefore, the coupling strength $g_{0}$ can be enhanced to be larger than the decay rate $\gamma_{a}$ of mode $a$ and even the resonance frequency $\Delta_{b}$ when we take $|\beta_{\textrm{ss}}|\gg 1$, and consequently the system can enter the ultrastrong coupling regime. (iii) The phase angle $\theta$ of the quadrature operator mode $b$ can be controlled by choosing proper driving phase in $\Omega_{b}$. This feature can be used to implement various geometric quantum operations such as the Kerr interaction and quantum gates.

The only approximation in the above derivation is the omission of the cross-Kerr interaction in the transformed Hamiltonian $H_{\textrm{tra}}$ in the regime of $\chi\ll \Delta_{b}$. To evaluate the adequacy of this approximation~\cite{seeSM}, we conduct numerical simulation of this system with the full Hamiltonian and the approximated Hamiltonian~(\ref{HapproxedMT}). To avoid the crosstalk of the dissipations on the approximation, we consider the closed system case~\cite{seeSM} by numerically solving the Schr\"{o}dinger equation. We then calculate the fidelity between the exact state $\vert \psi_{\textrm{ext}}(t)\rangle$ and the approximate state $\vert \psi_{\textrm{app}}(t)\rangle$ from the simulation of~(\ref{HapproxedMT}). We choose the initial state of the system as $\vert \psi(0)\rangle=|1\rangle_{a}|0\rangle_{b}$. In this case, the fidelity $F(t)=|\langle \psi_{\textrm{ext}}(t)\vert \psi_{\textrm{app}}(t)\rangle|$ can be obtained for the case of $\theta=0$ as $F(t)=\exp[-\Lambda(t)/2]$, with $\Lambda(t)=\left|g_{0}(1-e^{-i\Delta_{b}t})/\Delta_{b}-g_{0}(1-e^{-i(\Delta_{b}+\chi)t})/(\Delta_{b}+\chi)\right|^{2}$.

\begin{figure}[tbp]
\center
\includegraphics[bb=12 2 510 237, width=0.475 \textwidth]{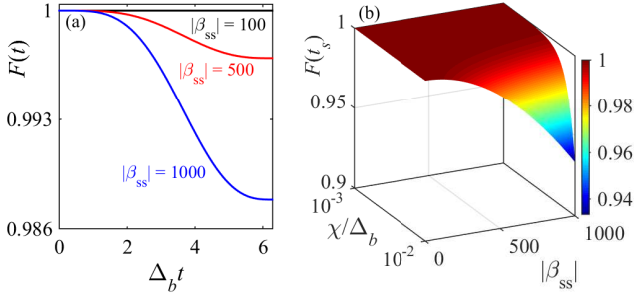}
\caption{(Color online) (a) The fidelity $F(t)$ as a function of the time $\Delta_{b}t$ when $\chi/\Delta_{b}=0.005$ and $|\beta_{\textrm{ss}}|=100$, $500$, and $1000$, which correspond to $g_{0}/\Delta_{b}=0.5$, $2.5$, and $5$, respectively. (b) The fidelity $F(t_{s})$ at time $t_{s}=\pi/\Delta_{b}$ as a function of the parameters $|\beta_{\textrm{ss}}|$ and $\chi/\Delta_{b}$.}
\label{Fig1}
\end{figure}

In Fig.~\ref{Fig1}(a), we show the fidelity $F(t)$ as a function of the time $t$ with $\chi/\Delta_{b}=0.005$ and $|\beta_{\text{ss}}|=100$, $500$, and $1000$. We see that the fidelity decreases for larger values of $|\beta_{\text{ss}}|$. This can be explained from the expression of $F(t)$ that the exponential decreasing rate is proportional to $|\beta_{\text{ss}}|^{2}$ in this case. Nevertheless, the fidelity can be very high because of $\chi\ll\Delta_{b}$. In Fig.~\ref{Fig1}(b), we display the fidelity $F(t_{s})$ at time $t_{s}=\pi/\Delta_{b}$ (the time for generation of cat state in mode $b$) as a function of $|\beta_{\text{ss}}|$ and $\chi/\Delta_{b}$. Here the fidelity is large in a wide parameter space and it is higher for smaller $\chi/\Delta_{b}$. For a given value of $\chi/\Delta_{b}$, $F$ is higher for a smaller value of $\vert\beta_{\text{ss}}\vert$. With the parameters for creating moderate displacement, for example $\chi|\beta_{\text{ss}}|=\Delta_{b}$, the fidelity could be larger than $0.99$.
Note that this fidelity is independent of $\omega_{a}^{\prime}$ because the term $\omega_{a}^{\prime}a^{\dagger}a$ commutates with other terms in the Hamiltonian.

\begin{figure}[tbp]
\center
\includegraphics[bb=30 4 636 296, width=0.47 \textwidth]{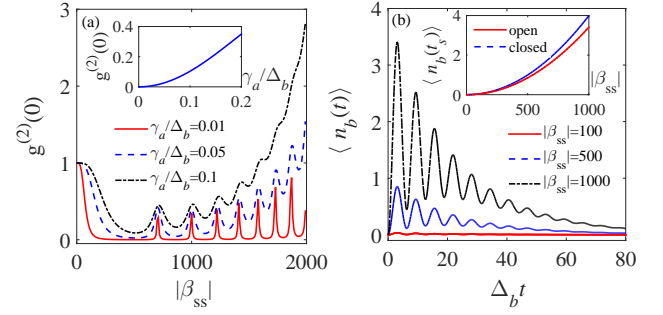}
\caption{(Color online) (a) The equal-time second-order correlation function $g^{(2)}(0)$ of mode $a$ as a function of $|\beta_{\textrm{ss}}|$ when $\gamma_{a}/\Delta_{b}=\gamma_{b}/\Delta_{b}=0.01$, $0.05$, and $0.1$ and under the single-photon resonance driving $\Delta'_{a}=g_{0}^{2}/(\Delta_{b}+\chi)$. Other parameters used in panel (a) are $\chi/\Delta_{b}=0.001$, $\bar{n}_{a}=\bar{n}_{b}=0$, and $\Omega_{a}/\gamma_{a}=0.1$. The inset shows the correlation function $g^{(2)}(0)$ at $g_{0}/\Delta_{b}=0.5$ as a function of $\gamma_{a}/\Delta_{b}$. (b) The dynamics of the average excitation number $\langle n_{b}(t)\rangle$ in mode $b$ when $\chi/\Delta_{b}=0.001$ and the enhanced factor takes various values $|\beta_{\textrm{ss}}|=100$, $500$, and $1000$. The inset shows the average excitation number $\langle n_{b}(t_{s})\rangle$ at time $t_{s}=\pi/\Delta_{b}$ as a function of $|\beta_{\textrm{ss}}|$ in both the closed- and open-system cases. Other parameters used in panel (b) are $\gamma_{a}/\Delta_{b}=\gamma_{b}/\Delta_{b}=0.05$ and $\bar{n}_{a}=\bar{n}_{b}=0$.}
\label{Fig2}
\end{figure}

\emph{Photon blockade}---One important application of the optomechanical interaction in the ultrastrong coupling regime is the photon blockade effect~\cite{Rabl2011}. The photon blockade effect can be seen from the dressed Kerr nonlinearity in the diagonalized Hamiltonian $V^{\dagger}H_{\textrm{app}}V=\omega_{a}^{\prime}a^{\dagger}a+\Delta_{b}b^{\dagger}b-(g_{0}^{2}/\Delta_{b})a^{\dagger}aa^{\dagger}a$,
with $V=\exp[(g_{0}/\Delta_{b})a^{\dagger}a(b^{\dagger}e^{i\theta}-be^{-i\theta})]$~\cite{seeSM}. To observe the photon blockade effect, the magnitude of the self-Kerr nonlinearity should be much larger than the decay rate, namely $g_{0}^{2}/\Delta_{b}\gg\gamma_{c}$, such that the anharmonicity in the energy levels can be resolved. In our scheme, the single-photon optomechanical coupling strength is enhanced by the large coherent displacement $|\beta_{\text{ss}}|$ and a small driving detuning $\Delta_{b}$. Here we should point out that the small detuning will not affect the thermal occupation number because $\bar{n}_{b}$ is determined by the natural frequency $\omega_{b}$ of mode $b$.
In Fig.~\ref{Fig2}(a), we plot the equal-time second-order correlation function $g^{(2)}(0)=\langle a^{\dag}a^{\dag}aa\rangle_{\text{ss}}/\langle a^{\dag}a\rangle_{\text{ss}}^{2}$ as a function of the enhanced factor $|\beta_{\text{ss}}|$ at various values of $\gamma_{a}/\Delta_{b}$. Here operator averages are for the steady state of the system~\cite{numericalcal}. We can see that the photon blockade effect (corresponding to $g^{(2)}(0)\ll1$) can be observed in the resolved-sideband limit $\gamma_{a}/\Delta_{b}\ll1$. The decay of mode $a$ will harm the photon blockade effect, as shown in the inset, where we display $g^{(2)}(0)$ as a function of $\gamma_{a}/\Delta_{b}$ at $g_{0}/\Delta_{b}=0.5$, which corresponds to the optimal $|\beta_{\text{ss}}|$ for photon blockade.

\emph{Macroscopic mechanical coherence}---Another important application of the optomechanical interaction in the ultrastrong coupling regime is the generation of the Schr\"{o}dinger cat state~\cite{Marshall2003} in mode $b$. The dynamical evolution of this system can be used to create the Schr\"{o}dinger cat states for both mode $a$~\cite{Mancini1997,Bose1997,Bose1999} and mode $b$~\cite{Marshall2003}. Up to the free evolution $\exp(-i\omega_{a}^{\prime}ta^{\dagger}a-i\Delta_{b}tb^{\dagger}b)$, the unitary evolution operator associated with the generalized optomechanical coupling $H_{\textrm{app}}$ can be written as $U_{\textrm{app}}(t)=\exp(i\lambda a^{\dagger}aa^{\dagger}a)D_{b}[(g_{0}/\Delta_{b})(e^{i\Delta_{b}t}-1)e^{i\theta}a^{\dagger }a]$,
where the Kerr parameter is given by $\lambda=(g_{0}^{2}/\Delta _{b}^{2})[\Delta_{b}t-\sin(\Delta_{b}t)]$ and $D_{b}[\beta]\equiv \exp(\beta b^{\dag}-\beta^{\ast}b)$ is the conditional displacement operator for mode $b$ with the conditional excitation number $a^{\dagger}a$ in mode $a$. At specific times $\Delta_{b}t=2n\pi/\Delta_{b}$  for natural numbers $n$, the two modes decouple and then the dynamics of mode $a$ corresponds to a Kerr interaction, which can used to create cat states. The conditional displacement for mode $b$ can also be used to create macroscopically distinct superposition. To this end, we consider an initial state of the system $|\Psi(0)\rangle=(1/\sqrt{2})(|0\rangle_{a}+|1\rangle_{a})|0\rangle_{b}$. The state at time $t$ of the system can be obtained as
$|\Psi(t)\rangle=(1/2)[|+\rangle_{a}(|0\rangle_{b}+e^{i\vartheta(t)}|\eta(t)\rangle_{b})+|-\rangle_{a}(|0\rangle_{b}-e^{i\vartheta(t)}|\eta(t)\rangle_{b})]$, where $|\pm\rangle_{a}=(|0\rangle_{a}\pm |1\rangle_{a})/\sqrt{2}$, $\vartheta(t)=(g^{2}_{0}/\Delta^{2}_{b})[\Delta_{b}t-\sin(\Delta_{b}t)]-\omega_{a}^{\prime}t$, and $\eta(t)=(g_{0}/\Delta_{b})(1-e^{-i \Delta_{b} t})$. The maximal displacement $\eta_{\textrm{max}}=2g_{0}/\Delta_{b}$ is obtained at $t=(2n+1)\pi/\Delta_{b}$ for natural numbers $n$. When $g_{0}/\Delta_{b}>1$, the coherent state $|\eta(t)\rangle_{b}$ can be approximately distinguished from the vacuum state $|0\rangle_{b}$, and hence macroscopically distinct superposed coherent states in mode $b$ can be generated by measuring mode $a$ in states $|\pm\rangle_{a}$. The single-photon displacement of mode $b$ can be seen by calculating the average excitation $\langle n_{b}(t)\rangle=\langle b^{\dag}b\rangle=|\eta(t)|^{2}$ in mode $b$. In Fig.~\ref{Fig2}(b), we show the dynamics of $\langle n_{b}(t)\rangle$ for several values of $|\beta_{\text{ss}}|$ in the presence of dissipations. The plots show that a larger maximal accessible displacement can be obtained for a larger $|\beta_{\text{ss}}|$, and that the dissipations will decrease the peak value of the displacement. In the inset, we plot the variable $\langle n_{b}(t_{s})\rangle$, which corresponds to the maximal displacement $\eta_{\text{max}}$, as a function of $|\beta_{\text{ss}}|$. We see that the maximal displacement could be larger than the zero-point fluctuation of mode $b$ (i.e., $\langle n_{b}\rangle>1$). This means that a quantum superposition of macroscopically distinct states in mode $b$ can be prepared with this method~\cite{seeSM}.

\emph{Geometrical quantum operations}---The generalized nonlinear interaction between the two modes $a$ and $b$ in Hamiltonian $H_{\textrm{app}}$ can be used to create a self-Kerr nonlinear interaction of mode $a$ via a sequence of operations. Consider the resonant driving case $\Delta_{b}=0$, and the corresponding unitary evolution operator becomes $U(t,\theta)=\exp(-i\omega_{a}^{\prime}ta^{\dagger}a)\exp[ig_{0}ta^{\dagger}a(b^{\dagger}e^{i\theta}+be^{-i\theta})]$,
which takes the form of an evolution operator associated with the conditional quadrature operator $X(\theta)=(b^{\dagger}e^{i\theta}+be^{-i\theta})/\sqrt{2}$. With the above unitary evolution operator, a self-Kerr interaction of mode $a$ can be obtained by designing a chain of unitary evolution based on the unconventional geometrical phase effect~\cite{Zhu2003} as
$U_{\textrm{tot}}=U(t,3\pi/2)U(t,\pi)U(t,\pi/2)U(t,0)=\exp(-4i\omega^{\prime}_{a}ta^{\dagger }a)\exp(2ig_{0}^{2}t^{2}a^{\dagger}aa^{\dagger}a)$. The unitary evolution operator $U_{\textrm{tot}}$ represents a pure self-Kerr interaction of mode $a$, and it is different from the transformed Kerr nonlinearity $U_{\textrm{app}}(t)$ associated with the optomechanical coupling $H_{\textrm{app}}$. The pure self-Kerr interaction is independent of the phonon states and hence the two modes are decoupled from each other with no phonon sidebands. However, in the optomechanical interactions, the eigenstates are the number state for mode $a$ dressed by the displaced number states for mode $b$~\cite{Liao2012}. Moreover, the phase shift associated with the Kerr interaction $2g_{0}^{2}t^{2}$ is continuously tunable and it can reach $\pi$ which is needed for realization of logic gate for quantum computation. The Kerr interaction in $U_{\textrm{app}}(t)$ only works at time $t=2n\pi/\Delta_{b}$.
\begin{figure}[tbp]
\center
\includegraphics[bb=60 29 720 315, width=0.47\textwidth]{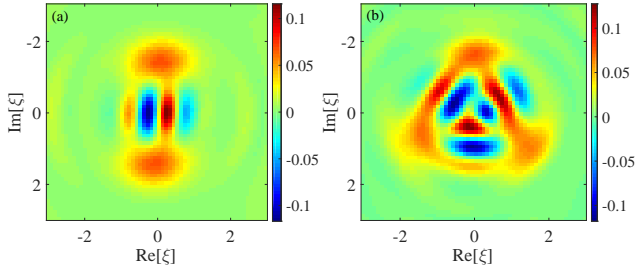}
\caption{(Color online) The Wigner functions of the generated states at $\tau=\pi$ and $\tau=2\pi/3$ in the presence of dissipation. The corresponding states in the idea case are the cat state $\psi(\tau=\pi)\rangle_{a}$ and the kitten state $\vert\psi(\tau=2\pi/3)\rangle_{a}$, respectively. Other parameters are given by $\alpha=2$, $\gamma_{a}/\Delta_{b}=\gamma_{b}/\Delta_{b}=0.05$, and $\bar{n}_{a}=\bar{n}_{b}=0$.}
\label{Fig3}
\end{figure}

The geometric Kerr interaction can also be used to create the Schr\"{o}dinger cat and kitten states~\cite{Birula1968,Miranowicz1990}. For simplicity, we express the unitary evolution operator as $U_{\textrm{tot}}=\exp[-i\phi(t)a^{\dagger}a]\exp[-i(\tau/2)a^{\dagger}a(a^{\dagger}a-1)]$, with $\phi(t)=4\omega^{\prime}_{a}t-2g^{2}_{0}t^{2}$ and $\tau=4g_{0}^{2}t^{2}$. For the initial state $|\psi(0)\rangle_{a}=|\alpha\rangle_{a}$, we consider the case of some specific  times $\tau=2\pi M/N$ with two coprime integers $M$ and $N$, the state of mode $a$ can be expressed as $|\psi(\tau)\rangle_{a}=\sum_{k=0}^{2N-1}c_{k}|\alpha e^{i\varphi_{k}}\rangle_{a}$, where $\varphi_{k}=2\pi k/(2N)$ and the coefficients are given by $c_{k}=(1/2N)\sum_{k=0}^{2N-1}\exp\{-i\frac{\pi}{N}[kn-Mn(n-1)]\}$. Here we omit the free evolution $\exp[-i\phi(t=\sqrt{M\pi/(2Ng^{2}_{0})})a^{\dagger}a]$ because this operator corresponds to a whole rotation of the state in phase space. For example, we choose $N=2$ and $M=1$, the state becomes $\vert \psi(\tau=\pi)\rangle_{a}=(1/\sqrt{2})(e^{-i\pi/4}\vert i\alpha\rangle_{a}+e^{i\pi/4}\vert-i\alpha\rangle_{a})$. When $N=3$ and $M=1$, we obtain a kitten state with three superposition components as $\vert\psi(\tau=2\pi/3)\rangle_{a}=(1/\sqrt{3})(e^{i\pi/6}\vert \alpha\rangle_{a}-i\vert\alpha e^{i2\pi/3}\rangle_{a}+e^{i\pi/6}\vert \alpha e^{i4\pi/3}\rangle_{a})$. In Fig.~\ref{Fig3} we plot the Wigner function $W(\xi)=\frac{2}{\pi}\text{Tr}[D_{a}^{\dagger}(\xi)\rho_{a}D_{a}(\xi)(-1)^{a^{\dagger}a}]$~\cite{Barnettbook} of the exact generated states, where $\xi$ is a complex variable, $\rho_{a}$ is the density matrix of the state, and $D_{a}(\xi)$ is the usual displacement operator for mode $a$. We can observe quantum interference pattern, which is a clear signature of quantum superposition~\cite{seeSM}.

\emph{Discussions}---Though we focus on the steady-state displacement in the above discussions, this method works for both the steady-state and the transient displacements $\beta_{\textrm{ss}}$ and $\beta(t)$. In the latter case, we can obtain a modulated optomechanical coupling $g(t)=\chi\beta(t)$. For example, we can choose a proper driving amplitude $\Omega_{b}(t)$ such that a sinusoidal enhancement $g_{0}\sin(\omega_{d}t)$ is obtained, where $\omega_{d}$ is the modulation frequency. It has been proved that the modulated optomechanical coupling can be used to enhance the photonic nonlinearity and to generate macroscopic superposition states~\cite{Liao2014}.

Our scheme can be implemented either by two electromagnetic field modes or by one electromagnetic mode and one mechanical mode coupled by the cross-kerr interaction~\cite{seeSM}. The requirements on the parameters are: $m\chi\ll\Delta_{b}$, $\chi|\beta_{\textrm{ss}}|\sim\Delta_{b}$, and $\Delta_{b}\gg\gamma_{a}$. We can choose proper driving frequency $\omega_{Lb}$ and amplitude $\Omega_{b}$ such that $\Delta_{b}\gg \gamma_{a}$ and  $\chi|\beta_{\textrm{ss}}|\sim\Delta_{b}$. For two cross-Kerr coupled microwave field modes, $\gamma_{a,b}/(2\pi)$ are of the order of $10^{4}$ - $10^{5}$ Hz~\cite{Aspelmeyer2014}. For one microwave field and one mechanical mode, the decay rates can be $\gamma_{a}/(2\pi)\sim10^{4}$ - $10^{5}$ Hz and $\gamma_{b}/(2\pi)\sim10^{2}$ - $10^{3}$ Hz~\cite{Aspelmeyer2014}. In these two cases, we choose $\Delta_{b}\sim10\gamma_{a}$. Corresponding to $\chi|\beta_{\textrm{ss}}|\sim\Delta_{b}$, we can choose $\chi/(2\pi)\sim10^{2}$ - $10^{3}$ Hz and $|\beta_{\textrm{ss}}|=10^{3}$. These parameters are accessible with current experimental technology. Note that the Kerr-type interactions in various quantum optical systems have been evaluated~\cite{Rebic2009,Hu2011,Nigg2012,Hoi2013,Holland2015,Bourassa2012,Majer2007,Thompson2008,Sankey2010,Karuza2013,Gong2009,Semiao2005,Maurer2004}.

\emph{Conclusions.}---We proposed a practical method to realize a generalized ultrastrong optomechanical coupling. This is achieved by driving one of the two bosonic modes coupled through a cross-Kerr interaction. We analyzed the parameter conditions under which this proposal works. We also studied the application of this scheme on the photon blockade effect, the cat state generation, and the implementation of geometric gates. This proposal provide a reliable method for studying few-photon optomechanics or simulating the optomechanical-type interactions between two electromagnetic fields with current experimental techniques.

\emph{Acknowledgments.}---J.-Q.L. is supported in part by NSFC Grant No.~11774087 and
HNNSFC Grant No.~2017JJ1021. J.-F.H. is supported by the NSFC Grant No.~11505055. L.T. is supported by the NSF (USA) under Award No. PHY-1720501. L.-M.K. is supported by the NSFC Grants No.~11375060, No.~11434011, and No.~11775075.
C.P.S. is supported by the National Basic Research Program of China Grants No.~2014CB921403 and No.~2016YFA0301201, the NSFC Grants No.~11421063 and No.~11534002, and the NSAF Grant No.~U1530401.

\newpage
\onecolumngrid
\newpage
\begin{center}
\textbf{\large Supplementary materials for ``Generalized Ultrastrong Optomechanics"}
\end{center}
\setcounter{equation}{0}
\setcounter{figure}{0}
\setcounter{table}{0}
\setcounter{page}{1}
\makeatletter
\renewcommand{\theequation}{S\arabic{equation}}
\renewcommand{\thefigure}{S\arabic{figure}}

This document consists of four parts: (I) Analyses of the parameter space of
the optomechanical model; (II) Derivation of the approximate Hamiltonian $H_{\text{app}}$
and evaluation of the parameter condition of the approximation;
(III) Detailed calculations of the applications of the generalized
optomechanical coupling; (IV) Discussions on the experimental implementation.

\section{I. Analyses of the parameter space of the optomechanical model}

In this section, we present some analyses on the parameter space of a standard cavity optomechanical system driven by a monochromatic field. First of all, we want to point out that the notations in this section are independent of the notations used in the main text and other sections in this supplemental material. This is because the motivation of this section is just to discuss the parameter space of a typical optomechanical system without other additional interaction terms.

For a typical optomechanical model, it is formed by a single-mode cavity field coupled to a single-mode mechanical oscillation via a radiation-pressure interaction (i.e., the optomechanical coupling). In order to manipulate this coupled system, a monochromatic laser field
is usually introduced to drive the cavity field. The Hamiltonian of this system reads
\begin{equation}
H=\omega_{a}a^{\dagger}a+\omega_{b}b^{\dagger}b-g_{0}a^{\dagger}a(b^{\dagger}+b)+(\Omega_{a}a^{\dag}e^{-i\omega_{d}t}+\Omega_{a}^{\ast}ae^{i\omega_{d}t}),  \label{Hoptdrivn}
\end{equation}
where $a$ ($a^{\dagger}$) and $b$ ($b^{\dagger}$) are the annihilation (creation) operators of the cavity field and the mechanical mode, respectively, with the corresponding resonance frequencies $\omega_{a}$ and $\omega_{b}$. The parameter $g_{0}$ is the single-photon optomechanical-coupling strength between the cavity field and the mechanical mode.  The parameters $\omega_{d}$ and $\Omega_{a}$ are the driven frequency and driving amplitude, respectively. By performing a rotating transformation with respect to $\omega_{d}a^{\dag}a$, the time factor in Hamiltonian~(\ref{Hoptdrivn}) can be eliminated and then the Hamiltonian becomes
\begin{equation}
H_{I}=\Delta_{a}a^{\dagger}a+\omega_{b}b^{\dagger}b-g_{0}a^{\dagger}a(b^{\dagger}+b)+(\Omega_{a}a^{\dag}+\Omega_{a}^{\ast}a),  \label{Hopttimeindep}
\end{equation}
where the driving detuning $\Delta_{a}=\omega_{a}-\omega_{d}$ is introduced.

To include the dissipations in this system, we assume that the cavity mode is coupled to a vacuum bath and the mechanical mode is coupled to a heat bath at a finite temperature.
In this case, the evolution of the optomechanical system is governed by the quantum master equation
\begin{eqnarray}
\dot{\rho}=i[\rho,H_{I}]+\gamma_{a}\mathcal{D}[a]\rho+\gamma_{b}(\bar{n}_{b}+1)\mathcal{D}[b]\rho+\gamma_{b}\bar{n}_{b}\mathcal{D}[b^{\dag}]\rho,\label{mastereqori}  \label{Hopttimeindep}
\end{eqnarray}
where $\mathcal{D}[o]\rho=o\rho o^{\dagger}-(o^{\dagger}o\rho+\rho o^{\dagger}o)/2$ is the standard Lindblad superoperator for bosonic-mode damping. The parameters $\gamma_{a}$ and $\gamma_{b}$ are the damping rates of the cavity mode and the mechanical mode, respectively. The parameter $\bar{n}_{b}$ is the thermal excitation occupation number of mode $b$'s heat bath.

In a typical open cavity optomechanical system, the relating parameters can be listed as:
\begin{eqnarray}
\omega_{a}&\rightarrow& \textrm{cavity field resonance frequency},\nonumber\\
\omega_{b}&\rightarrow& \textrm{mechanical mode resonance frequency},\nonumber\\
g_{0}&\rightarrow& \textrm{single-photon optomechanical-coupling strength},\nonumber\\
\Omega_{a}&\rightarrow& \textrm{cavity field driving magnitude}, \textrm{tunable parameter},\nonumber\\
\Delta_{a}&=&\omega_{a}-\omega_{d}\rightarrow\textrm{cavity field driving detuning}, \textrm{tunable parameter through the driving frequency $\omega_{d}$},\nonumber\\
\gamma_{a}&\rightarrow& \textrm{cavity field decay rate},\nonumber\\
\gamma_{b}&\rightarrow& \textrm{mechanical mode deday rate},\nonumber\\
\bar{n}_{b}&\rightarrow& \textrm{thermal excitation occupation number of the mechanical environment}.
\end{eqnarray}

Below we will analyze the relationship among these parameters. In this system, the cavity frequency is usually sufficient large such that the thermal excitation occupation number in the cavity's bath is negligible. From the view point of energy level transition, the cavity driving detuning $\Delta_{a}$ is a more important parameter to affect the dynamics of the system, and $\Delta_{a}$ is a tunable parameter via changing the driving frequency $\omega_{d}$. The mechanical frequency $\omega_{b}$ is an important parameter in this system because the ratio $\omega_{b}/\gamma_{a}>1$ is the sideband-resolution condition. This condition decides if the phonon sidebands can be resolved from the cavity emission spectrum. The single-photon optomechanical-coupling strength $g_{0}$ is also a very important parameter. This is because, on one hand, this ratio $g_{0}/\gamma_{a}$ is used to characterize the single-photon strong-coupling regime in optomechanics. Only when $g_{0}>\gamma_{a}$, the optomechanical phenomenon induced by a single photon can be observed in this system. On the other hand, the optomechanical coupling describes a constant force performed on the mechanical resonator and hence this is an unresonant interaction with a detuning $\omega_{b}$. In order to create quantum superposition of macroscopically distinct states, the relation $g_{0}/\omega_{b}>1$ should be satisfied. In few-photon optomechanics, the involved photon number is small and hence the driving magnitude $\Omega_{a}$ should be much smaller than the cavity-field decay rate, i.e., $\Omega_{a}/\gamma_{a}\ll1$. As described above, the cavity-field decay rate $\gamma_{a}$ is an important parameter because this quantity determines the condition for the sideband resolution and the single-photon strong coupling. In most optomechanical systems, the decay rate of the mechanical mode is very small. However, the thermal excitation number $\bar{n}_{b}$ in mode $b$ is an important parameter. This is because the thermal noise will prevent the observation of quantum effect in this system. As a result, the system should be cooled in advance to approach its ground state for observing quantum effects.
\begin{figure}[tbp]
\center
\includegraphics[bb=66 496 518 727, width=0.7\textwidth]{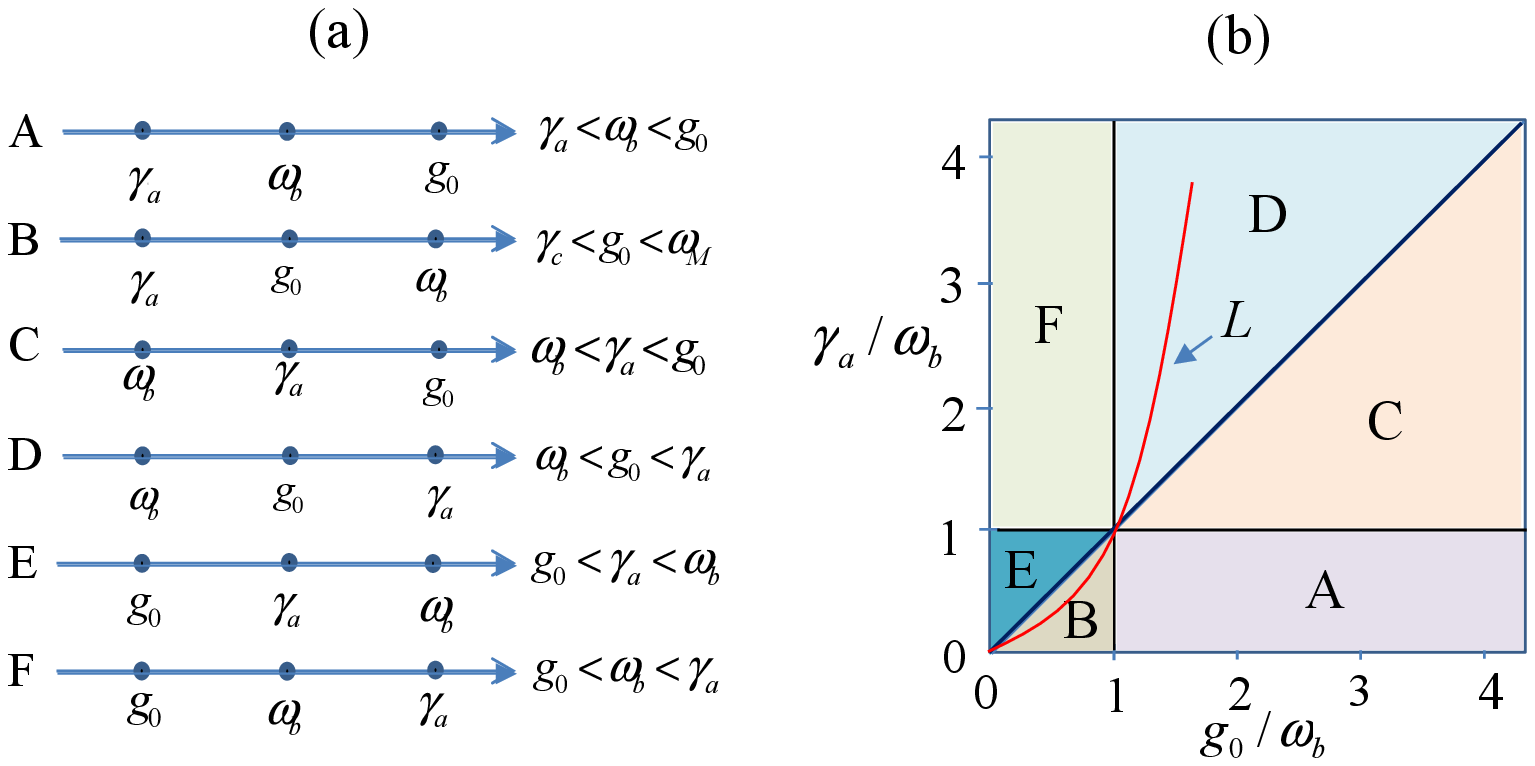}
\caption{(Color online) A parameter space diagram for cavity optomechanics in the weak-driving case. The parameter space is described by the single-photon optomechanical-coupling strength $g_{0}$, the photon decay rate $\gamma_{a}$, the resonance frequency $\omega_{b}$ of the mechanical resonator.}
\label{parameterspace}
\end{figure}

To analyze the parameter space, below we will focus on these three important parameters $g_{0}$, $\omega_{b}$, and $\gamma_{a}$. Usually, there are six cases of distribution for the three parameters, as shown in Fig.~\ref{parameterspace}(a). Relating to the single-photon optomechanical-coupling strength $g_{0}$, there exist three important parameter regimes: (i) The single-photon strong-coupling condition $g_{0}>\gamma_{a}$~\cite{Nunnenkamp2011,Rabl2011}, which is also related to the quantum parameter $g_{0}/\gamma_{a}>1$~\cite{Ludwig2008,Qian2012}, this condition guarantees that the cavity frequency shift caused by a zero-point fluctuation of the mechanical resonator can be resolved from the cavity spectrum. (ii) The strong dispersive coupling condition $g_{0}^{2}/\omega_{b}>\gamma_{a}$~\cite{Rabl2011}, which shows the condition for resolving the photonic Kerr nonlinear energy nonharmonicity from the cavity spectrum when $g_{0}\ll\omega_{b}$. (iii) The deep-strong coupling condition $g_{0}>\omega_{b}$, this condition depicts if the mechanical displacement forced by a single photon can be distinguished from the mechanical vacuum state. In addition, we should mention the resolved-sideband limit $\omega_{b}\gg\gamma_{a}$, which is not related to the coupling strength, but it is also very important in nonlinear optomechanics. This limit guarantees that the phonon sidebands can be well resolved in the cavity emission spectrum. Though the former three parameter regimes are very important in nonlinear optomechanics, they are not accessible by current experimental technologies. Based on the above analyses, we can introduce Fig.~\ref{parameterspace}(b) to describe the parameter space of an optomechanical system. The line $\gamma_{a}/\omega_{b}=1$ characterizes the sideband resolution condition. In Fig.~\ref{parameterspace}(b), the line $g_{0}/\omega_{b}=1$ describes the condition for creation of quantum superposition of macroscopically distinct states. The diagonal line $\gamma_{a}/g_{0}=1$ confirms that if the system satisfies the single-photon strong-coupling condition. In addition, the curve $L$ is determined by the relation $g_{0}^{2}/\omega_{b}=\gamma_{a}$. We should emphasize that the parameter condition $g_{0}^{2}/\omega_{b}>\gamma_{a}$ for evaluation of the photon blockade effect only works in the case of $g_{0}\ll\omega_{b}$. This is because the phonon sideband states will participate the photon transitions in the optomechanical system. The photon blockade effect is not monotonously stronger for a larger value of $g_{0}$. In particular, at the photon sideband resonance $g_{0}/\omega_{b}=\sqrt{m/2}$ for positive integers $m$, there is no photon blockade effect because the second photon transition is resonant~\cite{Liao2013}. In typical few-photon optomechanics, there are two important tasks. One is realization of the photon blockade effect, and the other is the generation of mechanical cat states. For observing photon blockade in optomechanics, the two conditions $\omega_{b}\gg \gamma_{a}$ and $g_{0}>\gamma _{a}$ (also $g_{0}^{2}/\omega_{b}>\gamma_{a}$) should be satisfied. It means that region $A$ and part of region $B$ are ok for observing photon blockade effect. To create a macroscopic mechanical cat state, the conditions $g_{0}>\omega_{b}$ and $\omega_{b}>\gamma_{a}$ should be satisfied. Therefore, only the region $A$ is ok. Note the the condition $\omega_{b}>\gamma_{a}$ guarantees that the maximal displacement has been created before the single photon emits out of the cavity.

\section{II. Derivation of the approximate Hamiltonian $H_{\text{app}}$
and evaluation of the parameter condition of the approximation}

In this section, we present a detailed derivation of the approximate Hamiltonian $H_{\text{app}}$. Hereafter, the notations are consistent with those used in the main text. We consider two bosonic modes $a$ and $b$ coupled via the cross-Kerr interaction. One of the two modes (for instance mode $b$) is driven by a monochromatic field. The Hamiltonian of the system reads
\begin{equation}
H=\omega_{a}a^{\dagger}a+\omega_{b}b^{\dagger}b+\chi a^{\dagger}ab^{\dagger}b
+(\Omega_{b}b^{\dagger}e^{-i\omega_{Lb}t}+\Omega_{b}^{\ast}be^{i\omega_{Lb}t}),
\end{equation}
where $a$ ($a^{\dagger}$) and $b$ ($b^{\dagger}$) are the annihilation (creation) operators of the two bosonic modes, with the corresponding resonance frequencies $\omega_{a}$ and $\omega_{b}$. The parameter $\chi$ is the coupling strength of the cross-Kerr interaction between the two modes. The mode $b$ is driven by a monochromatic field, with $\omega_{Lb}$ and $\Omega_{b}$ being the driving frequency and amplitude, respectively.

In a rotating frame with respect to $H_{0}=\omega_{Lb}b^{\dagger}b$, the Hamiltonian becomes
\begin{equation}
H_{I}=\omega_{a}a^{\dagger}a+\Delta_{b}b^{\dagger}b+\chi a^{\dagger}ab^{\dagger}b+(\Omega_{b}b^{\dagger}+\Omega_{b}^{\ast}b),
\end{equation}
where we introduce the driving detuning $\Delta_{b}=\omega_{b}-\omega_{Lb}$.

In the presence of dissipations, the evolution of the system is governed by the quantum master equation
\begin{eqnarray}
\dot{\rho}&=&i[\rho,H_{I}] +\frac{\gamma_{a}}{2}(\bar{n}_{a}+1)(2a\rho a^{\dagger}-a^{\dagger}a\rho-\rho a^{\dagger}a)
+\frac{\gamma_{a}}{2}\bar{n}_{a}(2a^{\dagger }\rho a-aa^{\dagger}\rho-\rho aa^{\dagger})\nonumber \\
&&+\frac{\gamma_{b}}{2}(\bar{n}_{b}+1)(2b\rho b^{\dagger}-b^{\dagger}b\rho-\rho b^{\dagger}b)
+\frac{\gamma_{b}}{2}\bar{n}_{b}(2b^{\dagger}\rho b-bb^{\dagger}\rho-\rho bb^{\dagger}),
\end{eqnarray}
where $\gamma_{a}$ ($\gamma_{b}$) and $\bar{n}_{a}$ ($\bar{n}_{b}$) are the damping rate and environment thermal excitation occupation number of mode $a$ ($b$), respectively.
In the strong-driving regime, the excitation number in mode $b$ is large and then mode $b$ contains a coherent part. This coherent part can be seen by performing the following displacement transformation
\begin{equation}
\rho^{\prime}=D_{b}(\beta)\rho D_{b}^{\dagger}(\beta),
\end{equation}
where $\rho^{\prime}$ is the density matrix of the two-mode system in the displacement representation, $D_{b}(\beta)=\exp(\beta b^{\dagger}-\beta^{\ast}b)$ is the displacement operator, and $\beta$ is the coherent displacement amplitude, which needs to be determined in the transformed master equation. Under the displacement transformation, the left-hand side of the master equation becomes
\begin{equation}
\dot{\rho}=\frac{d}{dt}[D_{b}^{\dagger}(\beta) \rho^{\prime }D_{b}(\beta)]
=\dot{D}_{b}^{\dagger}(\beta) \rho^{\prime}D_{b}(\beta)+D_{b}^{\dagger}(\beta) \dot{\rho}^{\prime }D_{b}(\beta)+D_{b}^{\dagger}(\beta)\rho^{\prime}\dot{D}_{b}(\beta).
\end{equation}
In terms of the relations
\begin{eqnarray}
\frac{d}{dt}D_{b}(\beta) &=&\dot{D}_{b}(\beta)=\frac{1}{2}(\dot{\beta}\beta ^{\ast }-\beta \dot{\beta}^{\ast})D_{b}(\beta)+D_{b}(\beta)(\dot{\beta}b^{\dagger }-\dot{\beta}^{\ast }b), \nonumber \\
\frac{d}{dt}D_{b}^{\dagger}(\beta) &=&\dot{D}_{b}^{\dagger}(\beta) =-\frac{1}{2}( \dot{\beta}\beta^{\ast}-\beta\dot{\beta}^{\ast}) D_{b}^{\dagger }(\beta)+(\dot{\beta}^{\ast}b-\dot{\beta}b^{\dagger}) D_{b}^{\dagger}(\beta),
\end{eqnarray}
the left-hand side of the master equation can be calculated as
\begin{eqnarray}
\dot{\rho}=\frac{d}{dt}[D_{b}^{\dagger}(\beta)\rho^{\prime}D_{b}(\beta)]=D_{b}^{\dagger}(\beta)\dot{\rho}^{\prime }D_{b}(\beta) +[D_{b}^{\dagger}(\beta) \rho ^{\prime}D_{b}(\beta),( \dot{\beta}b^{\dagger }-\dot{\beta}^{\ast}b)].
\end{eqnarray}
Using the relations
\begin{eqnarray}
D_{b}(\beta)bD_{b}^{\dagger}(\beta)&=&b-\beta,\hspace{1 cm}
D_{b}(\beta)b^{\dagger}D_{b}^{\dagger}(\beta)=b^{\dagger}-\beta^{\ast},
\end{eqnarray}
we proceed to derive the transformed master equation as
\begin{eqnarray}
\dot{\rho}^{\prime}&=&i[\rho^{\prime},(\omega_{a}+\chi\beta^{\ast}\beta)a^{\dagger}a+\Delta_{b}b^{\dagger}b
+\chi a^{\dagger}ab^{\dagger}b-\chi\beta a^{\dagger}ab^{\dagger}-\chi\beta^{\ast}a^{\dagger}ab]  \nonumber \\
&&+[\dot{\beta}+(i\Delta_{b}+\gamma_{b}/2)\beta -i\Omega_{b}](b^{\dagger}\rho^{\prime}-\rho^{\prime}b^{\dagger})
+[\dot{\beta}^{\ast}+(-i\Delta_{b}+\gamma_{b}/2)\beta^{\ast }+i\Omega _{b}^{\ast }](\rho^{\prime}b-b\rho^{\prime})  \nonumber \\
&&+\frac{\gamma_{a}}{2}(\bar{n}_{a}+1)(2a\rho^{\prime}a^{\dagger }-a^{\dagger }a\rho^{\prime }-\rho ^{\prime }a^{\dagger}a)
+\frac{\gamma_{a}}{2}\bar{n}_{a}(2a^{\dagger}\rho^{\prime}a-aa^{\dagger}\rho^{\prime}-\rho^{\prime }aa^{\dagger })\nonumber \\
&&+\frac{\gamma_{b}}{2}(\bar{n}_{b}+1)(2b\rho^{\prime}b^{\dagger }-b^{\dagger}b\rho^{\prime }-\rho ^{\prime }b^{\dagger}b)
+\frac{\gamma_{b}}{2}\bar{n}_{b}(2b^{\dagger}\rho^{\prime}b-bb^{\dagger}\rho ^{\prime}-\rho^{\prime }bb^{\dagger}).
\end{eqnarray}
The coherent part can be determined in the displacement representation when the coherent displacement amplitude $\beta$ obeys the equation
\begin{eqnarray}
\dot{\beta}+\left(i\Delta_{b}+\frac{\gamma_{b}}{2}\right)\beta-i\Omega_{b}=0.\label{betaEQ}
\end{eqnarray}
Then the quantum master equation in the displacement representation becomes
\begin{eqnarray}
\dot{\rho}^{\prime}&=&i[\rho^{\prime},H_{\text{tra}}]+\frac{\gamma_{a}}{2}(\bar{n}_{a}+1)( 2a\rho^{\prime}a^{\dagger}-a^{\dagger}a\rho^{\prime}-\rho^{\prime }a^{\dagger}a)+\frac{\gamma_{a}}{2}\bar{n}_{a}(2a^{\dagger}\rho^{\prime}a-aa^{\dagger}\rho^{\prime}-\rho^{\prime}aa^{\dagger})\nonumber\\
&&+\frac{\gamma_{b}}{2}(\bar{n}_{b}+1)(2b\rho^{\prime}b^{\dagger}-b^{\dagger}b\rho^{\prime}-\rho^{\prime}b^{\dagger}b)
+\frac{\gamma_{b}}{2}\bar{n}_{b}(2b^{\dagger}\rho^{\prime}b-bb^{\dagger}\rho^{\prime}-\rho^{\prime}bb^{\dagger}),\label{mateqexact}
\end{eqnarray}
where the transformed Hamiltonian in the displacement representation becomes
\begin{equation}
H_{\text{tra}}=(\omega_{a}+\chi\beta^{\ast}\beta)a^{\dagger}a+\Delta_{b}b^{\dagger}b-\chi a^{\dagger}a(\beta b^{\dagger}+\beta^{\ast}b)+\chi a^{\dagger}ab^{\dagger }b.\label{Hamtrasformed}
\end{equation}

Based on the tasks, we consider two cases of displacement: the steady-state displacement and the transient displacement. In the former case, the steady-state displacement amplitude can be obtained as
\begin{equation}
\beta_{\text{ss}}=\frac{\Omega_{b}}{\Delta_{b}-i\gamma_{b}/2}.\label{betassolutn}
\end{equation}
It can be seen from Eq.~(\ref{betassolutn}) that the coherent displacement amplitude $\beta_{\text{ss}}$ is tunable by choosing proper parameter $\Omega_{b}$ and $\Delta_{b}$.
The value of $|\beta_{\text{ss}}|$ could be very large in the strong-driving case $\Omega_{b}\gg\{\Delta_{b},\gamma_{b}\}$. For the transient-solution case, the optomechanical coupling becomes a time-dependent interaction. In particular, the interaction strength $g_{0}(t)$ is tailorable because we can obtain a desired $\beta(t)$ by designing a proper driving amplitude $\Omega_{b}(t)$.

In this work, we mainly focus on the steady-state displacement case, in which the time scale of the system approaching to its steady state is much shorter than other evolution time scales. In this case, the Hamiltonian becomes
\begin{equation}
H_{\text{tra}}=\omega_{a}^{\prime}a^{\dagger}a+\Delta_{b}b^{\dagger}b-g_{0}a^{\dagger}a(b^{\dagger}e^{i\theta}+be^{-i\theta})+\chi a^{\dagger}ab^{\dagger}b.\label{Hamtrasformedss}
\end{equation}
where we introduced the normalized frequency $\omega_{a}^{\prime}=\omega_{a}+\chi\vert\beta_{\text{ss}}\vert^{2}$, the enhanced coupling strength
\begin{eqnarray}
g_{0}=\chi\vert\beta_{\text{ss}}\vert,
\end{eqnarray}
and the phase angle $\theta$ of the quadrature operator of mode $b$, which is defined by $\beta_{\text{ss}}=\vert\beta_{\text{ss}}\vert e^{i\theta}$.

The motivation of ultrastrong optomechanics is to study the few-photon physics in optomechanical system, then we focus the few-photon regime, and under the condition
\begin{equation}
\vert m\chi\vert \ll \Delta_{b},\label{approxcondition}
\end{equation}
with $m$ being the largest photon number involved in the system, we can neglect the cross-Kerr interaction term to obtain the approximate Hamiltonian as
\begin{eqnarray}
H_{\text{app}}=\omega_{a}^{\prime}a^{\dagger}a+\Delta_{b}b^{\dagger}b-g_{0}a^{\dagger}a(b^{\dagger}e^{i\theta}+be^{-i\theta}).\label{Hamapproxim}
\end{eqnarray}
This approximate Hamiltonian is the main result of this work. Here we can see that the effective frequency of mode $b$ is given by $\Delta_{b}$, and that the effective coupling strength of the generalized optomechanical coupling is given by $g_{0}\equiv\chi|\beta_{\text{ss}}|$. The effective frequency $\Delta_{b}$ is controllable by tuning the driving frequency $\omega_{Lb}$, and the generalized coupling strength could be largely enhanced to enter the ultrastrong coupling regime by choosing a proper driving amplitude $\Omega_{b}$. The form of the optomechanical coupling is generalized because this coupling takes the form of a product of the occupation number operator of mode $a$ and the quadrature operator of mode $b$. The quadrature angle $\theta$ can be tuned by choosing the driving frequency $\omega_{Lb}$ and amplitude $\Omega_{b}$.

In the derivation of the approximate Hamiltonian $H_{\text{app}}$, the only approximation is the omission of the cross-Kerr interaction term in the transformed Hamiltonian $H_{\text{tra}}$. The condition under which the approximation is justified is that the frequency shift of mode $b$ induced by the cross-Kerr interaction should be much smaller than its effective frequency $\Delta_{b}$ in the displacement representation. Blow, we evaluate the reasonability of this approximation by calculating the fidelity between the exact state and the approximate state. To avoid the crosstalk from the dissipations, we first consider the closed-system case, in which the evolutions of the exact state and the approximate state are governed by the exact Hamiltonian $H_{\text{tra}}$ and the approximate Hamiltonian $H_{\text{app}}$, respectively. We assume that the initial state of the system is $|\psi(0)\rangle=|m\rangle_{a}|0\rangle_{b}$ so that we can calculate the exact state and the approximate state analytically.

\begin{figure}[tbp]
\center
\includegraphics[bb=17 31 500 165, width=1\textwidth]{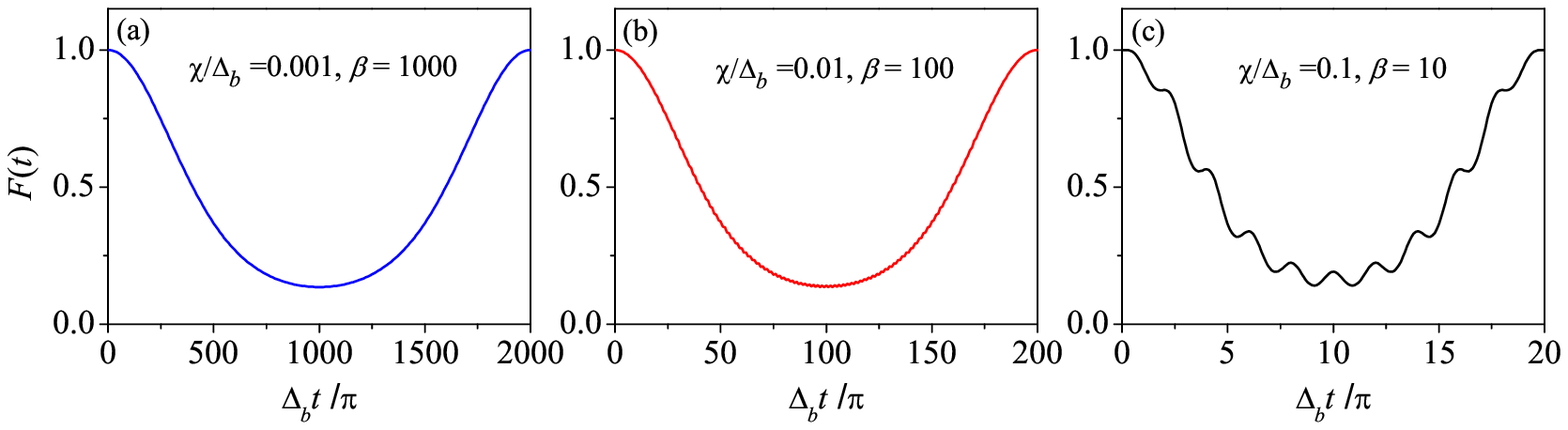}
\caption{(Color online) The fidelity defined by Eq.~(\ref{Fidvstclosed}) as a function of the evolution time when the parameters $\chi$ and $\beta$ take different values: (a) $\chi=0.001$ and $\beta=1000$, (b) $\chi=0.01$ and $\beta=100$, and (c) $\chi=0.1$ and $\beta=10$. The initial state of the system is $|1\rangle_{a}|0\rangle_{b}$.}
\label{Fvst1photonvarichi}
\end{figure}
Based on the exact Hamiltonian~(\ref{Hamtrasformed}), the exact state of the system at time $t$ can be obtained as
\begin{eqnarray}
\vert\psi_{\text{ext}}(t)\rangle=e^{-im\omega_{a}^{\prime}t}e^{i\zeta(t)}\vert m\rangle_{a}\vert\eta(t)\rangle_{b},
\end{eqnarray}
where the phase and the displacement amplitude are defined by
\begin{equation}
\zeta(t)=\frac{m^{2}\chi^{2}\vert\beta_{\text{ss}}\vert^{2}}{(\Delta_{b}+m\chi)^{2}}[(\Delta_{b}+m\chi)t-\sin[(\Delta_{b}+m\chi)t]],
\end{equation}
and
\begin{equation}
\eta(t)=\frac{m\chi\beta_{\text{ss}}}{(\Delta_{b}+m\chi)}[1-e^{-i(\Delta_{b}+m\chi)t}].
\end{equation}
Similarly, the approximate state of the system at time $t$ can be obtained, in terms of the approximate Hamiltonian~(\ref{Hamapproxim}), as
\begin{eqnarray}
\vert\psi_{\text{app}}(t)\rangle=e^{-im\omega_{a}^{\prime}t}e^{i\zeta^{\prime}(t)}\vert m\rangle_{a}\vert\eta^{\prime}(t)\rangle_{b}.
\end{eqnarray}
The phase and the displacement amplitude in this case are defined by
\begin{equation}
\zeta^{\prime}(t)=\frac{m^{2}\chi^{2}\vert\beta_{\text{ss}}\vert^{2}}{\Delta_{b}^{2}}[\Delta_{b}t-\sin(\Delta_{b}t)],
\end{equation}
and
\begin{equation}
\eta^{\prime}(t)=\frac{m\chi\beta_{\text{ss}}}{\Delta_{b}}(1-e^{-i\Delta_{b}t}).
\end{equation}
The fidelity between the exact state $\vert\psi_{\text{ext}}(t)\rangle$ and the approximate state $\vert\psi_{\text{app}}(t)\rangle$ can be calculated as
\begin{eqnarray}
F(t)&=&\left\vert\langle\psi_{\text{app}}(t)\vert\psi_{\text{ext}}(t)\rangle\right\vert\nonumber\\
&=&\exp\left[-\frac{1}{2}\left\vert\frac{m\chi\beta_{\text{ss}}}{\Delta_{b}}(1-e^{-i\Delta_{b}t})
-\frac{m\chi\beta_{\text{ss}}}{(\Delta_{b}+m\chi)}(1-e^{-i(\Delta_{b}+m\chi)t})\right\vert^{2}\right].\label{Fidvstclosed}
\end{eqnarray}
The properties of the fidelity corresponding to the single-photon initial state (i.e., $m=1$) have been shown in the main text. We want to point out, the fidelity is very high at the time scale characterized by the resonance frequency $\Delta_{b}$ (i.e., the typical time scale for manipulation of the mechanical mode). In addition, we show the fidelity in Fig.~\ref{Fvst1photonvarichi} as a function the evolution time. We can see that the fidelity will experience an approximate periodic oscillation between zero and one, with the time period $2\pi/\chi$, which is much longer than the time scale $\pi/\Delta_{b}$ for generation of mechanical cat states because of $\Delta_{b}\gg\chi$. For our purpose of state generation, the time scale is much shorter than the oscillation period, and hence the fidelity in our state generation scheme is very high. In addition, for a small value of $\chi/\Delta_{b}$, the fidelity experiences an approximate sinusoidal oscillation. With the increase of $\chi/\Delta_{b}$, the fidelity has some slight deviation of the sinusoidal function. In the presence of dissipation, the relaxation time is of the order of $1/\gamma_{a}$ and $1/\gamma_{b}$. The relaxation time is also much shorter than the time scale $2\pi/\chi$ because of $\gamma_{a,b}\gg\chi$, which shows that our scheme works in the weak-coupling regime of the cross-Kerr interaction.
\begin{figure}[tbp]
\center
\includegraphics[bb=6 36 468 300, width=1\textwidth]{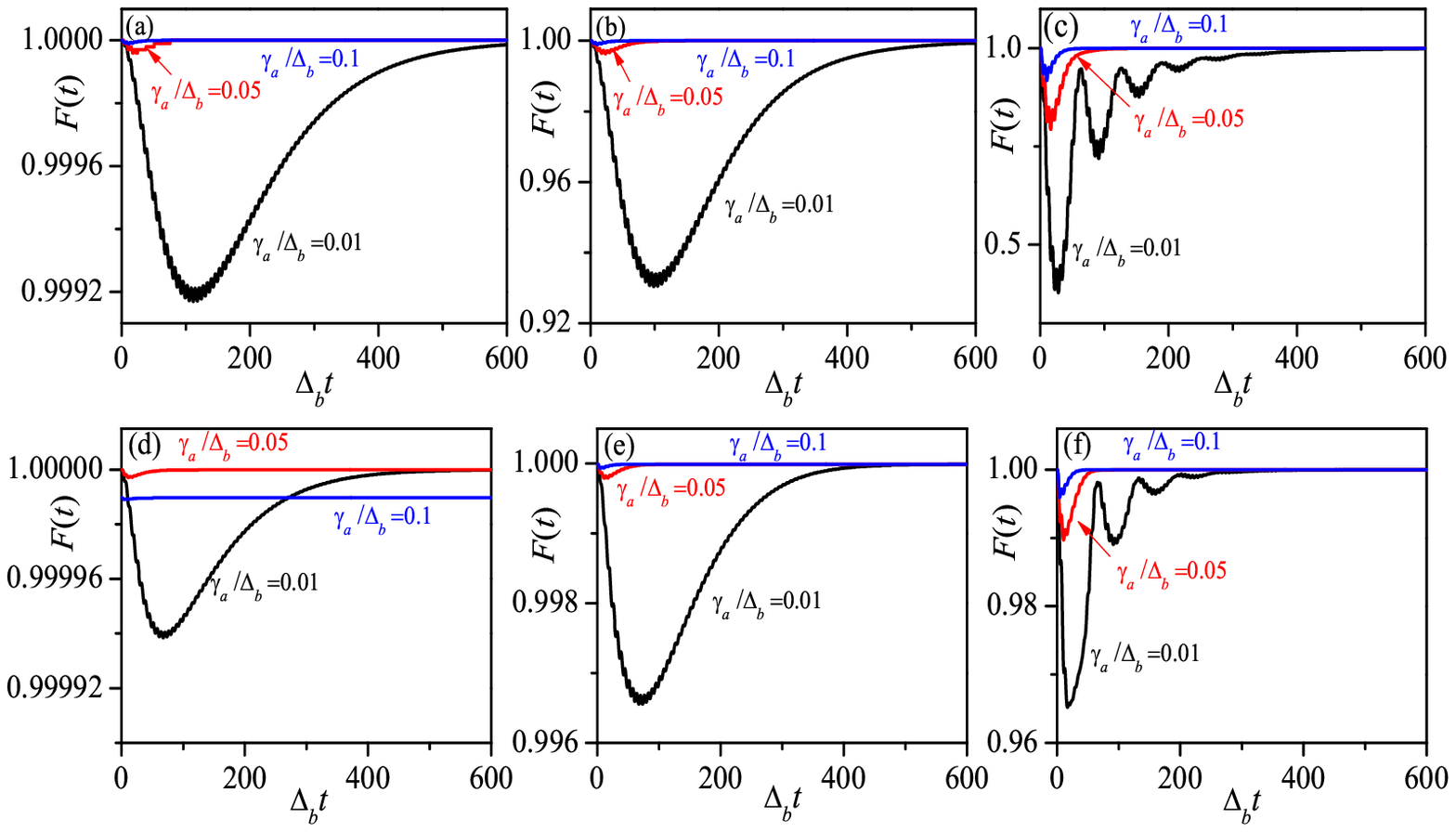}
\caption{(Color online) The dynamics of the fidelity defined by Eq.~(\ref{fidelityopen}) between the exact state and the approximate state in the open system case, when the parameter $\chi$ takes different values: (a) and (d) $\chi/\Delta_{b}=0.001$, (b) and (e) $\chi/\Delta_{b}=0.01$, and (c) and (f) $\chi/\Delta_{b}=0.1$. Here, the parameter $|\beta_{\text{ss}}|$ is chosen such that $\chi|\beta_{\text{ss}}|=\Delta_{b}$. We choose the initial state of the two modes as (a)-(c) $|1\rangle_{a}|0\rangle_{b}$ and (d)-(f) $|\alpha_{0}\rangle_{a}|\beta_{0}\rangle_{b}$ with $\alpha_{0}=\beta_{0}=0.2$. We take $\omega'_{a}=0$ because the fidelity is independent of this variable $\omega'_{a}$. Other parameters are given by $\bar{n}_{a}=\bar{n}_{b}=0$.}
\label{FvstopenSM}
\end{figure}

Below, we investigate the fidelity in the presence of dissipations. In this case, the evolution of the exact state and the approximate state is, respectively, governed by the exact master equation~(\ref{mateqexact}) and the approximation master equation, which takes the same form as Eq.~(\ref{mateqexact}) under the replacement $H_{\text{tra}}\rightarrow H_{\text{app}}$. To unify describe the equations of motion for the density matrix elements, we introduce a parameter $\eta$ into the equation of motion by the replacement $\chi\rightarrow \eta\chi$ only in the last term in Eq.~(\ref{Hamtrasformed}). The values of  $\eta=1$ and  $\eta=0$ correspond to the exact solution case and the approximate solution case, respectively. By expressing the density matrix of the two-mode system in the number-state representation as
\begin{equation}
\rho^{\prime}=\sum_{m,j,n,k=0}^{\infty}\rho^{\prime}_{m,j,n,k}\vert m\rangle_{a}\vert j\rangle_{b}\;_{a}\langle n\vert_{b}\!\langle k\vert,\hspace{1 cm}\rho_{m,j,n,k}^{\prime}=\;_{a}\!\langle m\vert_{b}\!\langle j\vert\rho^{\prime}\vert n\rangle_{a}\vert k\rangle_{b},
\end{equation}
we obtain the equations of motion for the density matrix elements as
\begin{eqnarray}
\dot{\rho}_{m,j,n,k}^{\prime}&=&\left\{i(n-m)\omega_{a}^{\prime}+i(k-j)\Delta_{b}+i\eta(nk-mj)\chi\right.\nonumber\\
&&\left.-\frac{\gamma_{a}}{2}[(m+n)(2\bar{n}_{a}+1)+2\bar{n}_{a}]-\frac{\gamma_{b}}{2}[(j+k)(2\bar{n}_{b}+1) +2\bar{n}_{b}]\right\}\rho_{m,j,n,k}^{\prime}\nonumber\\
&&-in\sqrt{k+1}\chi\beta\rho_{m,j,n,k+1}^{\prime}-in\sqrt{k}\chi\beta^{\ast}\rho_{m,j,n,k-1}^{\prime}+im\sqrt{j+1}\chi \beta^{\ast}\rho
_{m,j+1,n,k}^{\prime}+im\sqrt{j}\chi\beta\rho_{m,j-1,n,k}^{\prime}\nonumber\\
&&+\gamma_{a}(\bar{n}_{a}+1)\sqrt{(m+1)(n+1)}\rho_{m+1,j,n+1,k}^{\prime}+\gamma_{a}\bar{n}_{a}\sqrt{mn}\rho_{m-1,j,n-1,k}^{\prime}\nonumber\\
&&+\gamma_{b}(\bar{n}_{b}+1)\sqrt{(j+1)(k+1)}\rho_{m,j+1,n,k+1}^{\prime}+\gamma_{b}\bar{n}_{b}\sqrt{jk}\rho_{m,j-1,n,k-1}^{\prime}.
\end{eqnarray}
Based on the initial conditions, we can solve the equations of motion for these density matrix elements. Without loss of generality, in the simulations we assume that the initial state of the system is $\vert\alpha\rangle_{a}\otimes\vert\beta\rangle_{b}$, where $\vert\alpha\rangle$ and $\vert\beta\rangle$ are coherent states,  then we have
\begin{equation}
\rho^{\prime}_{m,j,n,k}(0)=e^{-\vert\alpha\vert^{2}}e^{-\vert\beta\vert^{2}}\frac{\alpha^{m}\alpha^{\ast n}\beta^{j}\beta^{\ast k}}{\sqrt{m!j!n!k!}}.
\end{equation}
We denote the density matrix corresponding to the two cases of $\eta=1$ and $\eta=0$
as $\rho_{\text{ext}}$ and $\rho_{\text{app}}$, respectively, then the fidelity between the exact density matrix
$\rho_{\text{ext}}$ and the approximate density matrix $\rho_{\text{app}}$ can be calculated by
\begin{equation}
F=\text{Tr}\left[\sqrt{\sqrt{\rho_{\text{ext}}}\rho_{\text{app}}\sqrt{\rho_{\text{ext}}}}\right].\label{fidelityopen}
\end{equation}
In Fig.~\ref{FvstopenSM}, we plot the fidelity given by Eq.~(\ref{fidelityopen}) as a function of the evolution time in the open-system case. Here we choose the initial state of the system as either $|1\rangle_{a}|0\rangle_{b}$ or $|\alpha\rangle_{a}|\beta\rangle_{b}$. In addition, we choose the parameters as $\chi/\Delta_{b}=0.001$, $0.01$, and $0.1$. The value of the displacement amplitude $|\beta_{\text{ss}}|$ is chosen such that $\chi|\beta_{\text{ss}}|=\Delta_{b}$. We can see that the fidelity is larger for a smaller value of the ratio $\chi/\Delta_{b}$, which is in consistent with the analysis on the parameter condition of the approximation. Owing to the dissipation, the fidelity experiences some oscillation and then approaches gradually to a stationary value. For a given value of $\chi$, the fidelity approaches to its stationary value in a faster manner for a larger decay rate.

\section{III. Detailed calculations of the applications of the generalized optomechanical coupling}

Below, we present the calculations of three applications of the generalized optomechanical coupling: the photon blockade effect in mode $a$, the generation of the Schr\"{o}dinger cat state in mode $b$, and the geometrically induced Kerr interaction and the generation of the Schr\"{o}dinger cat and kitten state.

\subsection{1. Photon blockade effect in mode $a$}

In this section, we study the photon blockade effect based on the generalized ultrastrong optomechanical coupling. For the Hamiltonian
\begin{equation}
H_{\text{app}}=\omega_{a}^{\prime}a^{\dagger}a+\Delta_{b}b^{\dagger}b-g_{0}a^{\dagger}a(b^{\dagger}e^{i\theta}+be^{-i\theta}),
\end{equation}
it can be diagonalized by the canonical transformation
\begin{eqnarray}
V=e^{\frac{g_{0}}{\Delta_{b}}a^{\dagger }a(b^{\dagger}e^{i\theta}-be^{-i\theta})}.
\end{eqnarray}
The transformed Hamiltonian becomes
\begin{eqnarray}
\tilde{H}_{\text{app}}=V^{\dagger}H_{\text{app}}V =\omega_{a}^{\prime}a^{\dagger}a+\Delta_{b}b^{\dagger }b-\frac{g_{0}^{2}}{\Delta_{b}}a^{\dagger}aa^{\dagger}a.
\end{eqnarray}
The effective Kerr parameter is given by $g_{0}^{2}/\Delta_{b}$, which is largely enhanced by either increasing the coupling strength $g_{0}$ or choosing a small driving detuning $\Delta_{b}$. We can choose proper parameters such that the Kerr parameter is much larger than the decay rate of mode $a$, i.e., $g_{0}^{2}/\Delta_{b}\gg\gamma_{a}$, then this Kerr effect can be used to realize the photon blockade effect.

To observe the photon blockade effect in mode $a$, we introduce a weak driving on mode $a$. Similar to the previous discussions, a strong driving is still performed on mode $b$ to  enhance the optomechanical coupling. Then the Hamiltonian of the system can be written as
\begin{equation}
H'=\omega_{a}a^{\dagger}a+\omega_{b}b^{\dagger}b+\chi a^{\dagger}ab^{\dagger}b+(\Omega_{a}a^{\dagger}e^{-i\omega_{La}t}+\Omega_{a}^{\ast}a^{\dagger}e^{i\omega_{La}t})+(\Omega_{b}b^{\dagger}e^{-i\omega_{Lb}t}+\Omega
_{b}^{\ast}be^{i\omega_{Lb}t}),
\end{equation}
Where $\Omega_{a}$ ($\Omega_{b}$) and $\omega_{La}$ ($\omega_{Lb}$) are the driving amplitude and frequency of mode $a$ ($b$). For observation of photon blockade, the driving field on mode $a$ is weak, i.e., $\Omega_{a}/\gamma_{a}\ll 1$. For enhancement of the optomechanical coupling, the driving of mode $b$ is strong. i.e., $\Omega_{b}/\gamma_{b}\gg 1$. Then we can treat the driving on mode $a$ as a perturbation in our calculations.

In a rotating frame with respect to
\begin{equation}
H'_{0}=\omega_{La}a^{\dagger}a+\omega_{Lb}b^{\dagger}b,
\end{equation}
the Hamiltonian becomes
\begin{equation}
H'_{I}=\Delta_{a}a^{\dagger}a+\Delta_{b}b^{\dagger}b+\chi a^{\dagger}ab^{\dagger}b
+(\Omega_{b}b^{\dagger}+\Omega_{b}^{\ast}b)+(\Omega_{a}a^{\dagger}+\Omega_{a}^{\ast}a^{\dagger})
\end{equation}
where we introduce the driving detunings
\begin{eqnarray}
\Delta_{a}=\omega_{a}-\omega_{La},\hspace{1 cm}\Delta_{b}=\omega_{b}-\omega_{Lb}.
\end{eqnarray}

In the open-system case, we add the same dissipation terms as the derivation in the above section. Since the driving on mode $a$ is weak and the driving on mode $b$ is strong, then we only perform the displacement transformation on mode $b$ as $\rho^{\prime}=D_{b}(\beta)\rho D_{b}^{\dagger}(\beta)$. Based on the fact that the displacement transformation operator commutates with the driving term of mode $a$, then the quantum master equation in the displacement representation reads
\begin{eqnarray}
\dot{\rho}^{\prime}&=&i\left[\rho^{\prime},H'_{\text{tra}}\right]\nonumber\\
&&+\frac{\gamma_{a}}{2}\left(\bar{n}_{a}+1\right)\left(2a\rho^{\prime}a^{\dagger}-a^{\dagger}a\rho^{\prime}-\rho^{\prime}a^{\dagger
}a\right)+\frac{\gamma_{a}}{2}\bar{n}_{a}\left(2a^{\dagger}\rho^{\prime}a-aa^{\dagger}\rho^{\prime}-\rho^{\prime}aa^{\dagger}\right)\nonumber\\
&&+\frac{\gamma_{b}}{2}\left(\bar{n}_{b}+1\right)\left(2b\rho^{\prime}b^{\dagger}-b^{\dagger}b\rho^{\prime}-\rho^{\prime}b^{\dagger
}b\right)+\frac{\gamma_{b}}{2}\bar{n}_{b}\left(2b^{\dagger}\rho^{\prime}b-bb^{\dagger}\rho^{\prime}-\rho^{\prime}bb^{\dagger}\right).\label{maseqphoblock}
\end{eqnarray}
Here the transformed Hamiltonian is given by
\begin{equation}
H'_{\text{tra}}=(\Delta_{a}+\chi\beta^{\ast}\beta)
a^{\dagger}a+\Delta_{b}b^{\dagger}b-\chi\beta a^{\dagger}ab^{\dagger}-\chi\beta^{\ast}a^{\dagger}ab+\chi a^{\dagger}ab^{\dagger}b+(\Omega_{a}a^{\dagger}+\Omega_{a}^{\ast}a^{\dagger}),
\end{equation}
where the displacement amplitude $\beta$ is determined by Eq.~(\ref{betaEQ}).

\begin{figure}[tbp]
\center
\includegraphics[bb= 5 3 443 402, width=0.55 \textwidth]{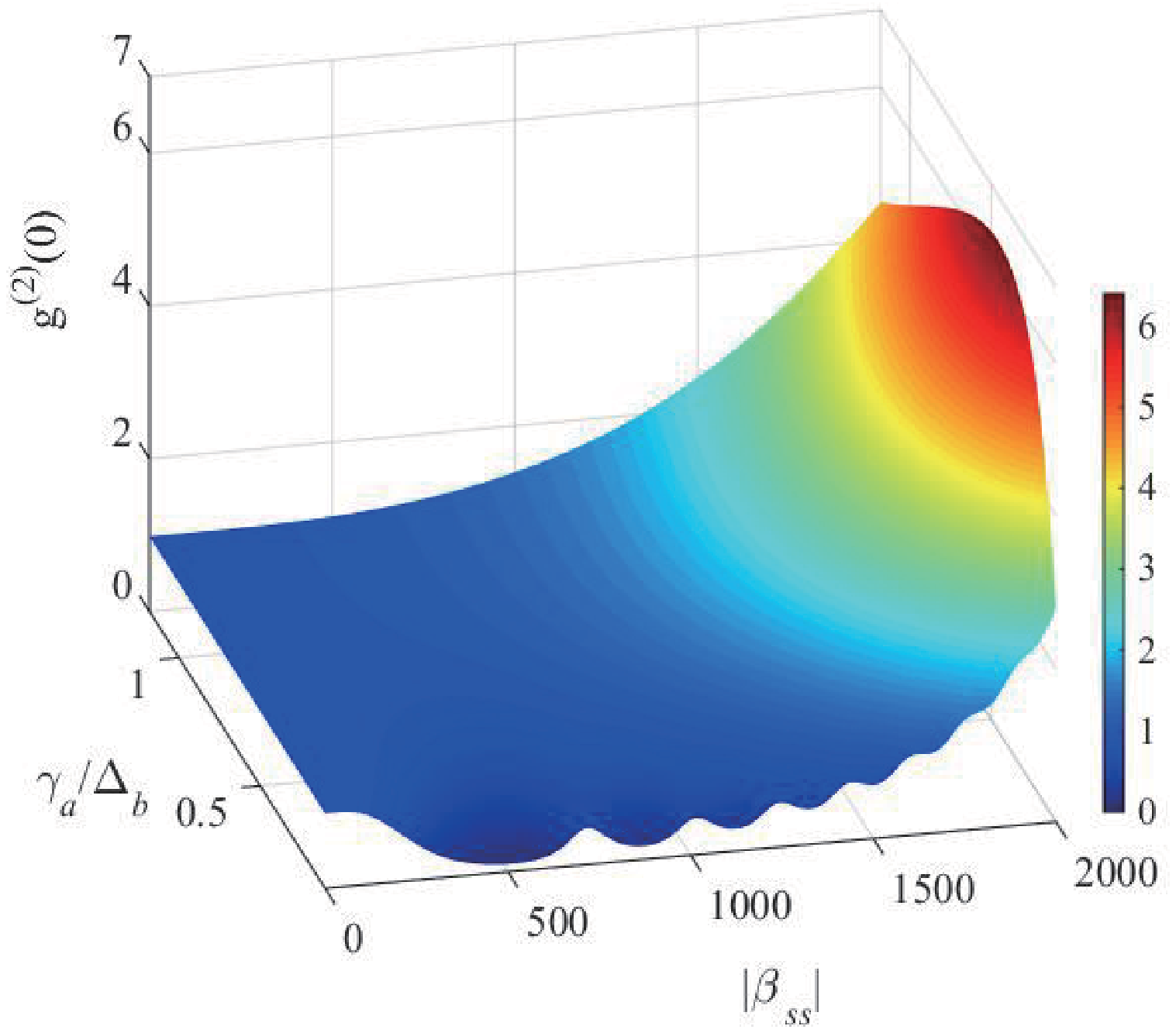}
\caption{(Color online) Plot of the equal-time second-order correlation function $g^{(2)}(0)$ as a function of the displacement amplitude $|\beta_{\text{ss}}|$ and the decay rate $\gamma_{a}/\Delta_{b}$. The driving detuning $\Delta_{a}=(\chi\beta_{\text{ss}})^{2}/(\Delta_{b}+\chi)$ is chosen such that the single-photon transition is resonant. Other parameters are given by $\chi/\Delta_{b}=0.001$, $\gamma_{b}=\gamma_{a}$, $\bar{n}_{a}=\bar{n}_{b}=0$, and $\Omega/\gamma_{a}=0.2$.}
\label{photonblockadeg2}
\end{figure}

Under the parameter condition~(\ref{approxcondition}) we can neglect the cross-Kerr interaction term to obtain the approximate Hamiltonian
\begin{eqnarray}
H'_{\text{app}}=\Delta_{a}^{\prime}a^{\dagger}a+\Delta_{b}b^{\dagger}b-g_{0}a^{\dagger}a(b^{\dagger}e^{i\theta}+be^{-i\theta})
+(\Omega_{a}a^{\dagger}+\Omega_{a}^{\ast}a^{\dagger}),
\end{eqnarray}
where we also consider the steady-state displacement case and then $\Delta_{a}^{\prime}=\Delta_{a}+\chi\vert\beta_{\text{ss}}\vert^{2}$ and $g_{0}=\chi\vert\beta_{\text{ss}}\vert$.
Below, we consider the case of $\theta=0$. Then the Hamiltonian becomes
\begin{equation}
H'_{\text{app}}=\Delta_{a}^{\prime}a^{\dagger}a+\Delta_{b}b^{\dagger}b-g_{0}a^{\dagger}a(b^{\dagger}+b)+\chi a^{\dagger}ab^{\dagger}b+\Omega_{a}(a^{\dagger}+a).\label{happphoblock}
\end{equation}
The undriven Hamiltonian of the system can be diagonalized as follows,
\begin{eqnarray}
&&\exp\left[-\frac{g_{0}a^{\dagger }a}{(\Delta_{b}+\chi a^{\dagger}a)}(b^{\dagger}-b)\right][
\Delta_{a}^{\prime }a^{\dagger}a+(\Delta_{b}+\chi a^{\dagger}a)b^{\dagger}b-g_{0}a^{\dagger}a(b^{\dagger}+b)]\exp\left[\frac{g_{0}a^{\dagger}a}{(\Delta_{b}+\chi
a^{\dagger}a)}(b^{\dagger}-b)\right]\nonumber \\
&=&\Delta_{a}^{\prime }a^{\dagger }a+(\Delta_{b}+\chi a^{\dagger}a)b^{\dagger}b-\frac{g_{0}^{2}(a^{\dagger}a)^{2}}{(\Delta_{b}+\chi a^{\dagger}a)}.
\end{eqnarray}
Then the eigensystem of the undriven Hamiltonian can be obtained as
\begin{eqnarray}
&&[\Delta_{a}^{\prime}a^{\dagger}a+(\Delta_{b}+\chi a^{\dagger}a)b^{\dagger}b-g_{0}a^{\dagger}a(b^{\dagger}+b)]\vert m\rangle_{a}\vert\tilde{j}(m)\rangle_{b}\nonumber\\
&=&\left[\Delta_{a}^{\prime }m+(\Delta_{b}+m\chi) j-\frac{g_{0}^{2}m^{2}}{(\Delta_{b}+\chi m)}\right]\vert m\rangle_{a}\vert\tilde{j}(m)\rangle_{b},\label{eigensystempb}
\end{eqnarray}
where the photon-number-dependent displaced Fock state of mode $b$ is defined by
\begin{equation}
\vert\tilde{j}(m)\rangle_{b}=\exp\left[\frac{g_{0}m}{(\Delta_{b}+\chi m)}(b^{\dagger}-b)\right]\vert j\rangle_{b}.
\end{equation}

Equation~(\ref{eigensystempb}) shows the photonic nonlinearity in the eigenstate energy spectrum, and this energy-level nonharmonicity is the physical origin of the appearance of photon blockade. Under the parameter condition $\chi\ll\Delta_{b}$, we can make the approximation $\Delta_{b}+m\chi\approx\Delta_{b}$, and then the present case is reduced to the approximate Hamiltonian case. In our numerical simulations, we solve the quantum master equation~(\ref{maseqphoblock}) numerically under the replacement of $H'_{\text{tra}}\rightarrow H'_{\text{app}}$ in Eq.~(\ref{happphoblock}). By calculating the equal-time second-order correlation function $g^{(2)}(0)$ in the steady state, we can evaluate the photon blockade effect in this system. To show the photon blockade effect, in Fig.~\ref{photonblockadeg2} we plot the correlation function $g^{(2)}(0)$ of mode $a$ as a function of the displacement amplitude $|\beta_{\text{ss}}|$ and the decay rate $\gamma_{a}/\Delta_{b}$. Here we choose the driving detuning $\Delta_{a}=(\chi\beta_{\text{ss}}|)^{2}/(\Delta_{b}+\chi)$ such that the first-photon transition is resonant. We can see that the correlation function $g^{(2)}(0)$ increases with the increase of the decay rate $\gamma_{a}$. For a small decay rate $\gamma_{a}$, the correlation function $g^{(2)}(0)$ exhibits some resonance peaks, which are induced by the phonon sideband resonant transitions. At some parameter spaces (the valley region) where the first photon transition is resonant and the second-photon transition is far-off-resonant, then mode $a$ will exhibit the photon blockade effect.

\begin{figure}[tbp]
\center
\includegraphics[bb=56 23 776 341, width=0.9 \textwidth]{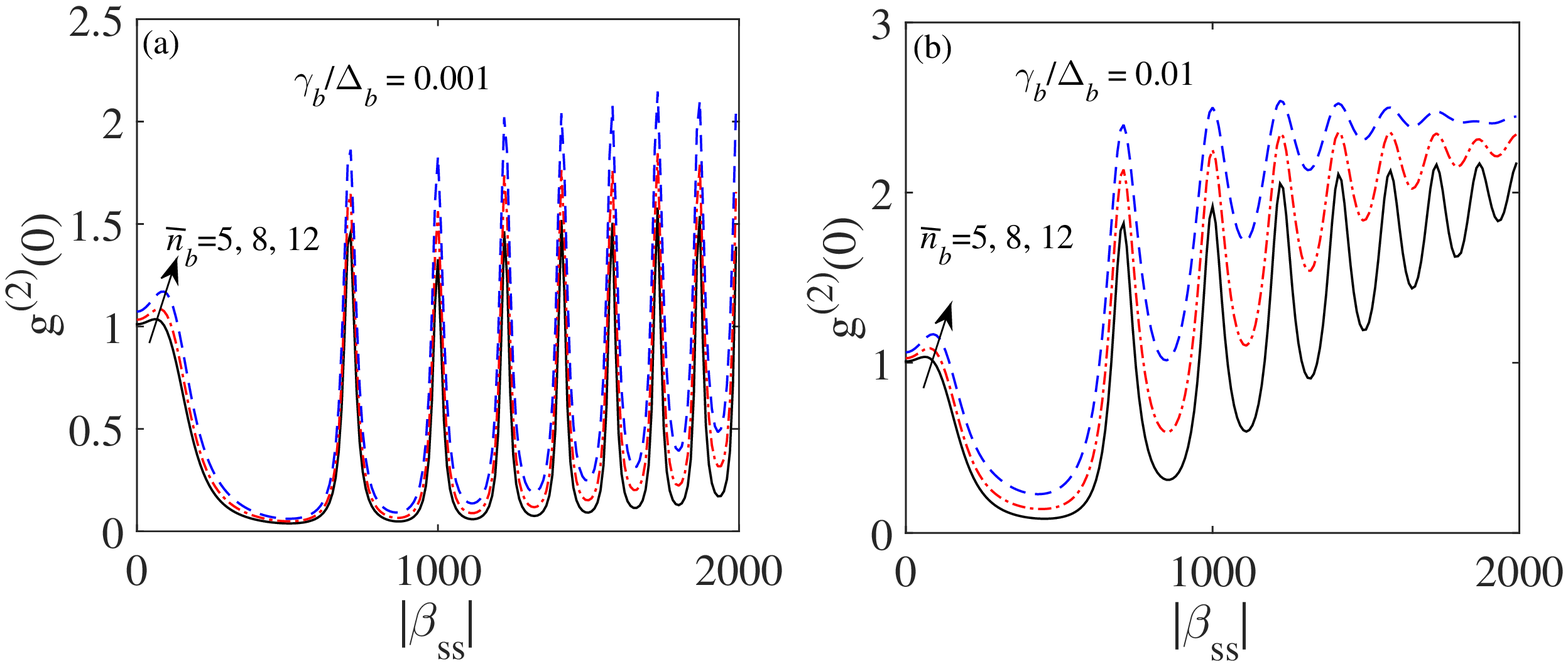}
\caption{(Color online) The equal-time second-order correlation function $g^{(2)}(0)$ of mode $a$ as a function of $|\beta_{\textrm{ss}}|$ when $\bar{n}_{a}=5$, $8$, and $12$. Here we choose different values of the decay rate of mode $b$: $\gamma_{b}/\Delta_{b}=0.001$ in panel (a) and $\gamma_{b}/\Delta_{b}=0.01$ in panel (b). The driving detuning of mode $a$ $\Delta'_{a}=g_{0}^{2}/(\Delta_{b}+\chi)$ is taken for satisfying the single-photon resonance condition. Other parameters used in are $\chi/\Delta_{b}=0.001$, $\gamma_{a}/\Delta_{b}=0.05$, $\bar{n}_{a}=0$, and $\Omega_{a}/\gamma_{a}=0.1$.}
\label{g2vsbetanbfiniteT}
\end{figure}

The thermal noise of the environment of mode $b$ will affect the photon blockade effect in mode $a$. This point can be seen by calculating the equal-time second-order correlation function $g^{(2)}(0)$. In Fig.~\ref{g2vsbetanbfiniteT}, we show the correlation function $g^{(2)}(0)$ as a function of $|\beta_{\text{ss}}|$ when the thermal occupation number takes different values $\bar{n}_{b}=5$, $8$, and $12$. Here we choose different values of the decay rate of mode $b$: $\gamma_{b}/\Delta_{b}=0.001$ in panel (a) and $\gamma_{b}/\Delta_{b}=0.01$ in panel (b). We can see that the lower envelop of the correlation function will increase with the increase of the thermal occupation number. This means that the thermal noise will harm the appearance of the photon blockade effect. For a small decay rate $\gamma_{b}/\Delta_{b}=0.001$, the photon blockade effect still exists for a moderately large thermal occupation number $\bar{n}_{b}$. For a larger value of $\gamma_{b}/\Delta_{b}=0.01$, the photon blockade effect in the large region of $|\beta_{\text{ss}}|$ will disappear gradually with the increase of the thermal occupation number $\bar{n}_{b}$.

\subsection{2. Generation of the Schr\"{o}dinger cat state in mode $b$}

In this section, we show the generation of the Schr\"{o}dinger cat state of mode $b$ in terms of the generalized optomechanical coupling.
We start our discussion by considering the Hamiltonian $H_{\text{app}}$ given by Eq.~(\ref{Hamapproxim}). In the rotating frame with respect to
\begin{equation}
H_{\text{app}}^{(0)}=\omega_{a}^{\prime}a^{\dagger}a+\Delta_{b}b^{\dagger}b,
\end{equation}
the Hamiltonian $H_{\text{app}}$ becomes
\begin{equation}
H_{\text{app}}^{(I)}(t)=-g_{0}a^{\dagger}a(b^{\dagger}e^{i\Delta_{b}t}e^{i\theta}+be^{-i\Delta_{b}t}e^{-i\theta}).
\end{equation}
The unitary evolution operator associated with the Hamiltonian $H_{\text{app}}$ can be
expressed as
\begin{equation}
U(t)=e^{-iH_{\text{app}}^{(0)}t}U_{\text{app}}(t),\label{totevounitopt}
\end{equation}
where $U_{\text{app}}(t)$ is the unitary evolution operator associated
with the Hamiltonian $H_{\text{app}}^{(I)}(t)$, its form is
\begin{equation}
U_{\text{app}}(t)=\mathcal{T}e^{-i\int_{0}^{t}H_{\text{app}}^{(I)}(t')dt^{\prime}},
\end{equation}
where ``$\mathcal{T}$" stands for the time-ordering integral.
The $U_{\text{app}}(t)$ is determined by the equation of motion
\begin{equation}
\dot{U}_{\text{app}}(t)=-iH_{\text{app}}^{(I)}(t)U_{\text{app}}(t),
\end{equation}
under the initial condition $U_{\text{app}}(0)=I$.
Using the Magnus proposal, the solution for $U_{\text{app}}(t)$ is a matrix exponential
\begin{equation}
U_{\text{app}}(t)=\exp[\Omega(t)],\hspace{1 cm}\Omega(t)=\sum_{k=1}^{\infty}\Omega_{k}(t),
\end{equation}
where $\Omega_{k}(t)$ are determined by the commutation between the Hamiltonian at different times.
Since the commutation of two Hamiltonian operators at different times $[H_{\text{app}}^{(I)}(t_{1}),H_{\text{app}}^{(I)}(t_{2})] $ is a c-number, so
there are no higher-order terms for $k>2$, namely $\Omega_{k>2}(t)=0$. In terms of the expression of $H_{\text{app}}^{(I)}(t) $, we have
\begin{eqnarray}
\Omega_{1}(t)&=&\int_{0}^{t}dt_{1}[-iH_{\text{app}}^{(I)}(t_{1})]=\frac{g_{0}}{\Delta_{b}}a^{\dagger}a[b^{\dagger}(e^{i\Delta_{b}t}-1) e^{i\theta}-b(e^{-i\Delta_{b}t}-1)e^{-i\theta}],
\end{eqnarray}
and
\begin{eqnarray}
\Omega_{2}(t)&=&\frac{1}{2}\int_{0}^{t}dt_{1}\int_{0}^{t_{1}}dt_{2}[-iH_{\text{app}}^{(I)}(t_{1}),-iH_{\text{app}}^{(I)}(t_{2})]=i\frac{g_{0}^{2}}{\Delta_{b}^{2}}[ \omega_{b}t-\sin(\Delta_{b}t)]a^{\dagger}aa^{\dagger}a,
\end{eqnarray}
then
\begin{eqnarray}
U_{\text{app}}(t)&=&\exp[\Omega_{1}(t)+\Omega_{2}(t)]  \nonumber \\
&=&\exp\left\{ i\frac{g_{0}^{2}}{\Delta _{b}^{2}}[\Delta_{b}t-\sin(\Delta_{b}t)]a^{\dagger}aa^{\dagger}a\right\}\exp\left\{\frac{g_{0}}{\Delta _{b}}a^{\dagger }a[b^{\dagger}(e^{i\Delta_{b}t}-1)e^{i\theta}-b(e^{-i\Delta _{b}t}-1)e^{-i\theta}]\right\}.\label{evoluoptcatstae}
\end{eqnarray}
We see clearly that this unitary evolution operator can be used to create the Schr\"{o}dinger cat states for mode $b$. To this end, we assume that the initial state of the system is
\begin{equation}
\vert\Psi(0)\rangle=\frac{1}{\sqrt{2}}(\vert 0\rangle_{a}+\vert 1\rangle_{a})\vert 0\rangle_{b}.\label{initstatecatmodeb}
\end{equation}
Then in terms of the unitary evolution operator~(\ref{totevounitopt}) we can calculate the state of the system at time $t$ as
\begin{eqnarray}
\vert\Psi(t)\rangle=\frac{1}{\sqrt{2}}[\vert 0\rangle_{a}\vert0\rangle_{b}+\exp[i\vartheta''(t)]\vert 1\rangle_{a}\vert\eta''(t)\rangle_{b}],
\end{eqnarray}
where the phase and displacement amplitude are defined by
\begin{eqnarray}
\vartheta''(t)&=&\frac{g_{0}^{2}}{\Delta_{b}^{2}}[\Delta_{b}t-\sin(\Delta_{b}t)]-\omega_{a}^{\prime}t,\nonumber \\
\eta''(t)&=&\frac{g_{0}}{\Delta_{b}}(1-e^{-i\Delta_{b}t})e^{i\theta}.
\end{eqnarray}
We see that the displacement reaches its maximum $\eta''_{\textrm{max}}=2g_{0}e^{i\theta}/\Delta_{b}$ at time $\Delta_{b}t=(2n+1)\pi$ for natural numbers $n$.

To create quantum superposition of mode $b$, we measure the state of mode $a$ with the bases
\begin{equation}
\vert\pm\rangle_{a}=(\vert0\rangle_{a}\pm\vert 1\rangle_{a})/\sqrt{2}.
\end{equation}
If we express the state of mode $a$ with the basis states $\vert\pm\rangle_{a}$, then the state of the system becomes
\begin{eqnarray}
\vert\Psi(t)\rangle&=&\frac{1}{2}\left[\vert+\rangle_{a}(\vert0\rangle_{b}+e^{i\vartheta''(t)}\vert\eta''(t)\rangle_{b})+\vert-\rangle_{a}(\vert 0\rangle_{b}-e^{i\vartheta''(t)}\vert\eta''(t)\rangle_{b})\right].
\end{eqnarray}
Corresponding to the states $\vert\pm\rangle_{a}$ are measured, mode $b$ collapses into the states
\begin{eqnarray}
|\phi_{\pm}(t)\rangle_{b}=\mathcal{N}_{\pm}(\vert0\rangle_{b}\pm e^{i\vartheta''(t)}\vert\eta''(t)\rangle_{b}),
\end{eqnarray}
where the normalization constants are defined by
\begin{equation}
\mathcal{N}_{\pm}=\left[2\left(1\pm e^{-\frac{\vert\eta''(t)\vert^{2}}{2}}\cos\vartheta''(t)\right)\right]^{-1/2}.
\end{equation}
The corresponding probabilities for the measured states $\vert\pm\rangle_{a}$ are
\begin{eqnarray}
P_{\pm}(t)=\frac{1}{2}\left[1+e^{-\frac{\vert\eta''(t)\vert^{2}}{2}}\cos\vartheta''(t)\right].
\end{eqnarray}
\begin{figure}[tbp]
\center
\includegraphics[bb=55 4 720 372, width=1 \textwidth]{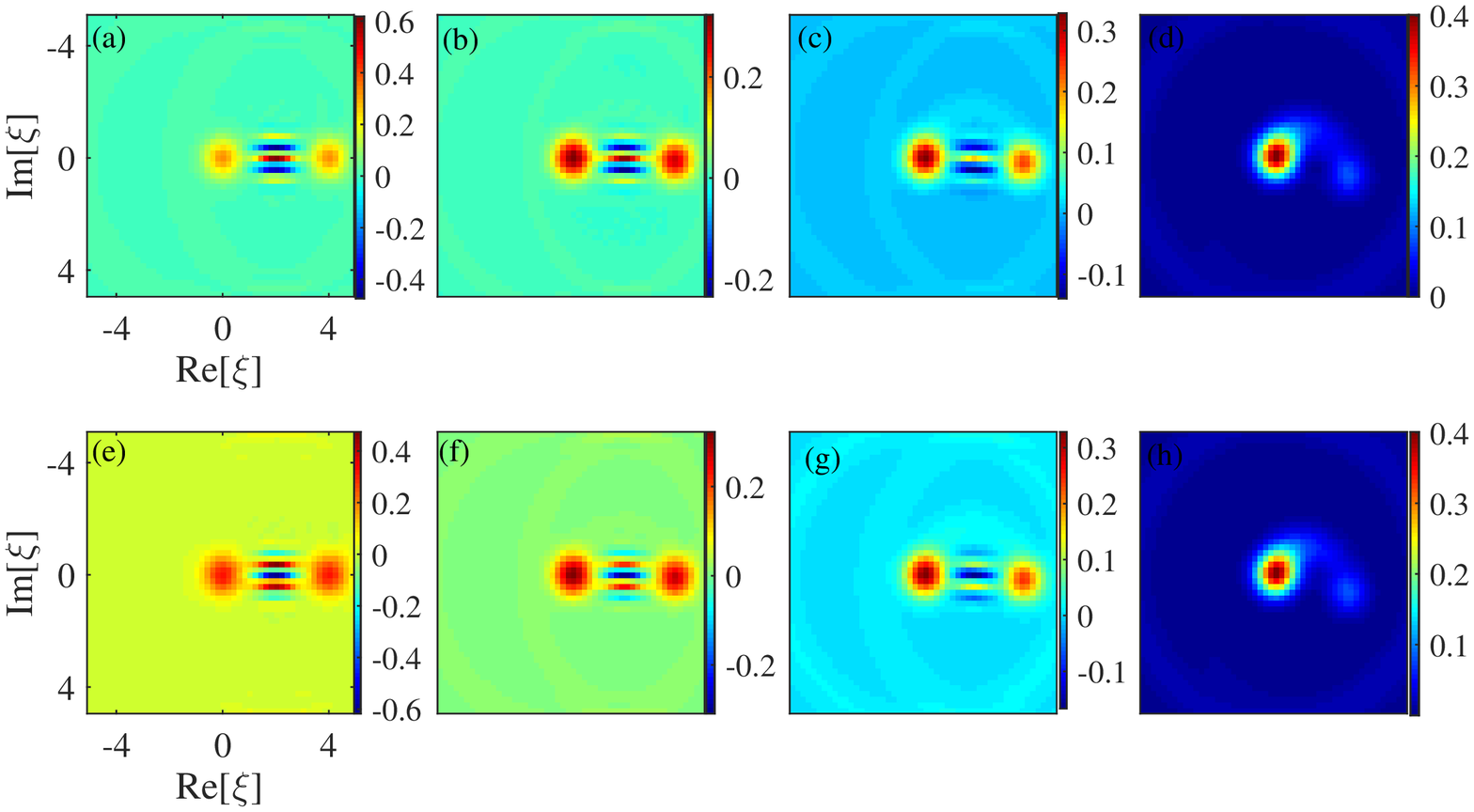}
\caption{(Color online) Plots the Wigner functions $W_{\rho_{b}^{(\pm)}}(\xi)$ defined by Eq.~(\ref{Wigfunccatstate}) of the generated state in mode $b$ when the decay rates take different values: (a) and (e) $\gamma_{a}/\Delta_{b}=\gamma_{b}/\Delta_{b}=0$, (b) and (f) $\gamma_{a}/\Delta_{b}=\gamma_{b}/\Delta_{b}=0.05$, (c) and (g) $\gamma_{a}/\Delta_{b}=\gamma_{b}/\Delta_{b}=0.1$, and (d) and (h) $\gamma_{a}/\Delta_{b}=\gamma_{b}/\Delta_{b}=0.5$. The Wigner functions in the first and second rows correspond to the states $\rho_{b}^{(+)}$ and $\rho_{b}^{(-)}$, respectively. Other parameters are given by $\chi/\Delta_{b}=0.001$, $\beta=2000$, $\bar{n}_{a}=\bar{n}_{b}=0$, and $t_{s}=\pi/(\Delta_{b}+\chi)$.}
\label{Wigfunmodebcat}
\end{figure}

In numerical simulations, we consider the transformed Hamiltonian $H_{\text{tra}}$ and include the dissipations of the two modes. By numerically solving the quantum master equations and performing the measurement at time $t_{s}=\pi/(\Delta_{b}+\chi)$, we can obtain two density matrices of mode $b$ corresponding to the two measurement states $|\pm\rangle$ of mode $a$. To solve this master equation, we denote the density matrix of the system at time $t$ as
\begin{eqnarray}
\rho(t)=\sum\limits_{m,j,n,k=0}^{\infty}\rho_{m,j,n,k}(t)\vert m\rangle_{a}\vert j\rangle_{b}\;_{a}\langle n\vert\;_{b}\langle k\vert.
\end{eqnarray}
By solving the equations of motion for the density matrix elements, we can obtain the density matrix $\rho(t)$. We proceed performing the measurement of mode $a$ into the states $\vert\pm\rangle_{a}$. After the measurement, the density matrices of mode $b$ become
\begin{eqnarray}
\rho_{b}^{(\pm)}(t_{s})=\mathcal{M}_{\pm}\sum\limits_{j,k=0}^{\infty}(\rho_{0,j,0,k}\pm\rho_{0,j,1,k}\pm\rho_{1,j,0,k}+\rho_{1,j,1,k})\vert j\rangle_{b}\;_{b}\langle k\vert,
\end{eqnarray}
where we introduce the variables
\begin{eqnarray}
\mathcal{M}_{\pm}=1/\sum\limits_{j=0}^{\infty}(\rho_{0,j,0,j}\pm\rho_{0,j,1,j}\pm\rho_{1,j,0,j}+\rho_{1,j,1,j}).
\end{eqnarray}

The Wigner function of the two density matrices can be calculated with the formula
\begin{eqnarray}
W(\xi)=\frac{2}{\pi}\text{Tr}[D_{b}^{\dagger}(\xi)\rho_{b}D_{b}(\xi)(-1)^{b^{\dagger}b}],
\end{eqnarray}
where $\xi$ is a complex variable, $\rho_{b}$ is the density matrix of mode $b$, and $D_{b}(\xi)=\exp(\xi b^{\dagger}-\xi^{\star}b)$ is the usual displacement operator for mode $b$.
The Wigner functions for the density matrices $\rho_{b}^{(\pm)}$ can be calculated as
\begin{eqnarray}
W_{\rho_{b}^{(\pm)}}(\xi)=\frac{2}{\pi}M_{\pm}\sum_{l,j,k=0}^{\infty}(\rho_{0,j,0,k}\pm\rho_{0,j,1,k}\pm\rho
_{1,j,0,k}+\rho_{1,j,1,k})(-1)^{l}\;_{b}\langle l\vert D_{b}(-\xi)\vert j\rangle
_{b}\;_{b}\langle k\vert D_{b}(\xi)\vert l\rangle_{b},\label{Wigfunccatstate}
\end{eqnarray}
where the matrix elements of the displacement operator in the Fock space can be calculated by~\cite{Buek1990}
\begin{eqnarray}
_{b}\!\langle m\vert D_{b}(\beta)\vert n\rangle\!_{b}=\left\{\begin{array}{c}\sqrt{\frac{m!}{n!}}e^{-\vert\beta\vert^{2}/2}(-\beta^{\ast})^{n-m}L_{m}^{n-m}(\vert\beta\vert^{2}),\hspace{0.3 cm}n>m, \\
\sqrt{\frac{n!}{m!}}e^{-\vert\beta\vert^{2}/2}(\beta)^{m-n}L_{n}^{m-n}(\vert\beta\vert^{2}),\hspace{0.3 cm}m>n,
\end{array}\right.\label{matrixelemD}
\end{eqnarray}
with $L_{n}^{m}(x)$ being the associated Laguerre polynomials.

\begin{figure}[tbp]
\center
\includegraphics[bb=47 14 544 285, width=1 \textwidth]{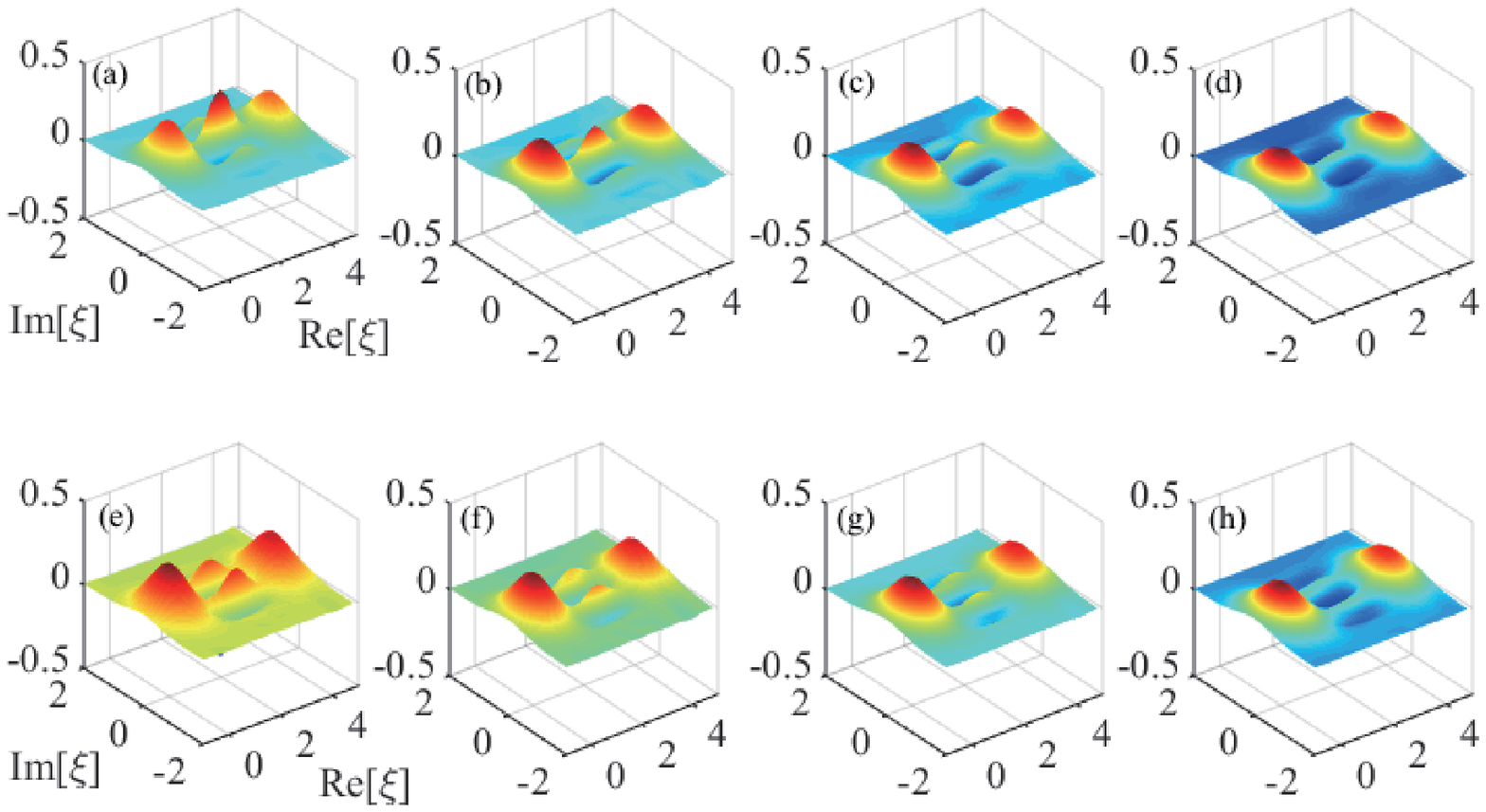}
\caption{(Color online) Plots the Wigner functions $W_{\rho_{b}^{(\pm)}}(\xi)$ defined by Eq.~(\ref{Wigfunccatstate}) of the generated state in mode $b$ when the thermal excitation number take different values: (a) and (e) $\bar{n}_{b}=1$, (b) and (f) $\bar{n}_{b}=3$, (c) and (g) $\bar{n}_{b}=5$, and (d) and (h) $\bar{n}_{b}=8$. The Wigner functions in the first and second rows correspond to states $\rho_{b}^{(+)}$ and $\rho_{b}^{(-)}$, respectively. Other parameters are given by $\chi/\Delta_{b}=0.001$, $\beta=2000$, $\gamma_{a}/\Delta_{b}=0.05$, $\gamma_{b}/\Delta_{b}=0.01$, $\bar{n}_{a}=0$, and $t_{s}=\pi/(\Delta_{b}+\chi)$.}
\label{WFmodebfiniteT}
\end{figure}

In Fig.~\ref{Wigfunmodebcat}, we plot the Wigner functions $W_{\rho_{b}^{(\pm)}}(\xi)$ of the generated states $\rho_{b}^{(\pm)}(t_{s})$ in mode $b$ when the decay rates take different values: (a) and (e) $\gamma_{a}/\Delta_{b}=0$, (b) and (f) $\gamma_{a}/\Delta_{b}=0.05$, (c) and (g) $\gamma_{a}/\Delta_{b}=0.1$, and (d) and (h) $\gamma_{a}/\Delta_{b}=0.5$. Panels (a)-(d) and (e)-(h) are plotted for the states $\rho_{b}^{(+)}$ and $\rho_{b}^{(-)}$, respectively. Here we can see that the Wigner functions exhibit clear evidence of macroscopically distinct superposition components and quantum interference pattern. With the increase of the decay rate, the interference pattern disappears gradually, and the main peak corresponding to the coherent component $|\eta''\rangle$ moves approaching to the origin, this is because the coherent state decays to the vacuum state in the presence of dissipation.

We also study the influence of the thermal occupation number in mode $b$ on the cat state generation. This consideration makes sense in the case where the mode $b$ is a mechanical resonator and hence the thermal occupation number cannot be neglected. Usually, for mechanical mode $b$, the decay rate $\gamma_{b}$ is smaller then that ($\gamma_{a}$) of the electromagnetic mode $a$. In Fig.~\ref{WFmodebfiniteT}, we plot the Wigner functions $W_{\rho_{b}^{(\pm)}}(\xi)$ of the states $\rho_{b}^{(\pm)}(t_{s})$ when the thermal occupation number takes different values: (a) and (e) $\bar{n}_{b}=1$, (b) and (f) $\bar{n}_{b}=3$, (c) and (g) $\bar{n}_{b}=5$, and (d) and (h) $\bar{n}_{b}=8$. The panels in the first and second rows correspond to states $\rho_{b}^{(+)}$ and $\rho_{b}^{(-)}$, respectively. We can see that the interference phenomenon (i.e., the oscillation between the main peaks) disappears gradually with the increase of the thermal occupation number $\bar{n}_{b}$.

\subsection{3. Geometrically induced Kerr interaction and Schr\"{o}dinger's cat and kitten state generation}

\subsubsection{(a) The closed-system case}

In this section, we study how to obtain a continuous self-Kerr interaction evolution of mode $a$ without the sidebands of mode $b$ by designing a
chain of unitary operation. To this end, we consider the resonant driving case $\Delta_{b}=0$ of mode $b$. In this case, the Hamiltonian becomes
\begin{equation}
H=\omega_{a}^{\prime}a^{\dagger}a-g_{0}a^{\dagger}a(b^{\dagger}e^{i\theta}+be^{-i\theta}).
\end{equation}
The unitary evolution operator associated with this resonant driving Hamiltonian is given by
\begin{equation}
U(t,\theta)=\exp[-i\omega_{a}^{\prime}ta^{\dagger}a]\exp[ig_{0}ta^{\dagger}a(b^{\dagger}e^{i\theta}+be^{-i\theta})].
\end{equation}
Here the phase angle $\theta$ is an important parameter to realize the geometric scheme. To implement a self-Kerr interaction for mode $a$, we need to use the following evolution sequence, which is formed by four steps of evolution,
\begin{eqnarray}
U(t,0)&=&\exp[-i\omega_{a}^{\prime}ta^{\dagger}a]\exp[ig_{0}ta^{\dagger}a(b^{\dagger}+b)],\nonumber \\
U(t,\pi/2)&=&\exp[-i\omega_{a}^{\prime}ta^{\dagger}a]\exp[-g_{0}ta^{\dagger}a(b^{\dagger}-b)],\nonumber \\
U(t,3\pi/2)&=&\exp[-i\omega_{a}^{\prime}ta^{\dagger}a]\exp[g_{0}ta^{\dagger}a(b^{\dagger}-b)],\nonumber \\
U(t,\pi)&=&\exp[-i\omega_{a}^{\prime}ta^{\dagger}a]\exp[-ig_{0}ta^{\dagger}a(b^{\dagger}+b)].
\end{eqnarray}
The total evolution operator can be calculated as
\begin{eqnarray}
U_{\text{tot}}&=&U(t,3\pi/2)U(t,\pi)U(t,\pi/2) U(t,0)\nonumber\\
&=&\exp[-4i\omega_{a}^{\prime}ta^{\dagger}a]\exp[2ig_{0}^{2}t^{2}a^{\dagger}aa^{\dagger}a]\nonumber\\
&=&\exp[-i\theta(t) a^{\dagger }a]\exp[2ig_{0}^{2}t^{2}(a^{\dagger }aa^{\dagger}a-a^{\dagger}a)],
\end{eqnarray}
with
\begin{equation}
\theta(t)=4\omega _{a}^{\prime}t-2g_{0}^{2}t^{2}.
\end{equation}
It is worth noting that this self-Kerr interaction is continuously accessible in the time domain, which is different from the dynamical evolution case, in which the self-Kerr interaction only accessible at the specific times [cf. Eq.~(\ref{evoluoptcatstae}), at time $\Delta_{b}t=2n\pi$ for positive integer number, the evolution operator is reduced to a self-Kerr interaction evolution].
\begin{figure}[tbp]
\center
\includegraphics[bb=24 10 536 264, width=0.55 \textwidth]{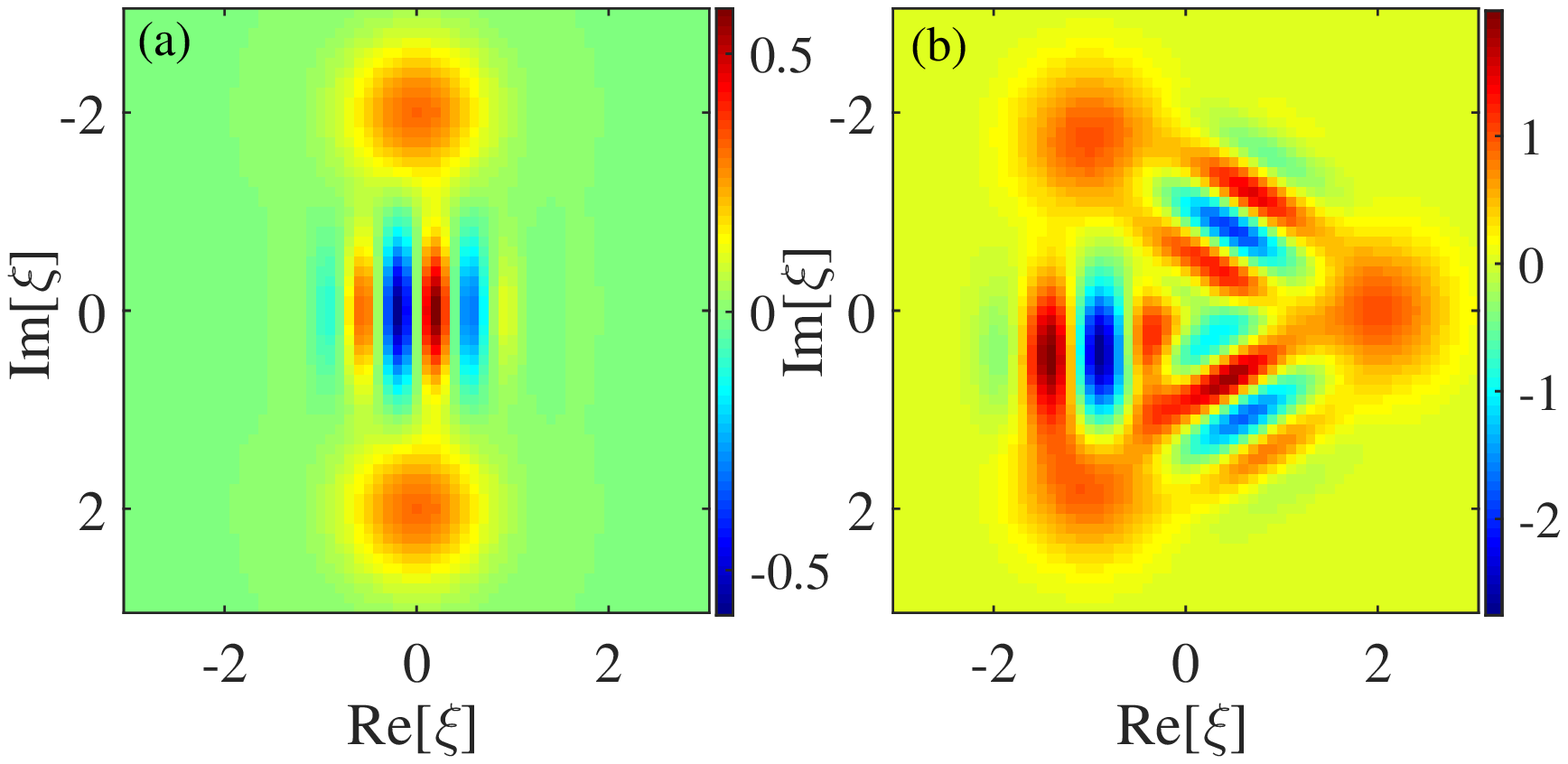}
\caption{(Color online) Plots of the Wigner functions, defined in Eqs.~(\ref{Wigfuncatstaanaly}) and~(\ref{Wigfunkittenstaanaly}), of the cat state $\psi(\tau=\pi)\rangle$ and the kitten state $\vert\psi(\tau=2\pi/3)\rangle$ defined in Eqs.~(\ref{catstageome}) and~(\ref{kittenstageome}), respectively. Here the coherent amplitude of the initial coherent state $|\alpha\rangle$ is $\alpha=2$.}
\label{Wignerfuncclose}
\end{figure}

To generate the Schr\"{o}dinger cat states, we assume that the initial state of mode $a$ is $\vert\alpha\rangle=e^{-\frac{\vert \alpha\vert ^{2}}{2}}\sum_{n=0}^{\infty }\frac{\alpha ^{n}}{\sqrt{n!}}\vert n\rangle$,
then, up to a free evolution $\exp(-4i\omega_{a}^{\prime}ta^{\dagger}a)$, the state of the system at time $t$ becomes
\begin{eqnarray}
\vert\psi(\tau)\rangle_{a}=\exp[2ig_{0}^{2}t^{2}(a^{\dagger }aa^{\dagger}a-a^{\dagger}a)]\vert\alpha\rangle_{a}=e^{-\frac{\vert \alpha\vert ^{2}}{2}}\sum_{n=0}^{\infty}
\frac{\alpha ^{n}}{\sqrt{n!}}\exp\left[i\frac{\tau }{2}n(n-1)\right]\vert n\rangle_{a},
\end{eqnarray}
where we introduce $\tau=4g_{0}^{2}t^{2}$. Since $n(n-1)$ is an even number, then we know that $\vert\Psi(\tau+2\pi)\rangle$ is a periodic function of $\tau$ with the period $T=2\pi$,
\begin{eqnarray}
\vert\psi(\tau+2\pi)\rangle_{a} &=&e^{-\frac{\vert\alpha\vert^{2}}{2}}\sum_{n=0}^{\infty}\frac{\alpha^{n}
}{\sqrt{n!}}\exp\left[i\frac{\tau+2\pi}{2}n(n-1)\right]\vert n\rangle_{a}=e^{-\frac{\vert\alpha\vert^{2}}{2}}\sum_{n=0}^{\infty}
\frac{\alpha^{n}}{\sqrt{n!}}e^{i\frac{\tau}{2}n(n-1)}\vert n\rangle_{a}.
\end{eqnarray}
Moreover, based on this relation
\begin{eqnarray}
e^{i\frac{\tau }{2}(n+2N)(n+2N-1)}=e^{i\frac{\tau }{2}n(n-1)}e^{i\tau N(2N+2n-1)},
\end{eqnarray}
we can see that if we choose the $\tau =\frac{M}{N}2\pi$, then we can express the state as
\begin{eqnarray}
\left\vert\psi\left(\frac{M}{N}T\right)\right\rangle_{a}=e^{-\frac{\vert\alpha\vert^{2}}{2}}\sum_{n=0}^{\infty}\frac{\alpha^{n}}{\sqrt{n!}}\exp\left[i\frac{M}{N}\pi n(n-1)\right]\vert n\rangle_{a}=\sum_{k=0}^{2N-1}c_{k}\vert\alpha e^{i\varphi_{k}}\rangle_{a},
\end{eqnarray}
where $\vert\alpha e^{i\varphi_{k}}\rangle_{a}$ are coherent states, with the phase $\varphi_{k}=\frac{k}{N}\pi$, $k=0,1,2,...,2N-1$.
The superposition coefficients are given by
\begin{equation}
c_{k}=\frac{1}{2N}\sum_{n=0}^{2N-1}e^{-i\frac{\pi}{N}[kn-Mn(n-1)]}.
\end{equation}
The evolution time is $4t$ with $t=\sqrt{\frac{\tau}{4g_{0}^{2}}}=\sqrt{\frac{M2\pi}{4g_{0}^{2}N}}$.

We now show two examples for the generation of the Schr\"{o}dinger cat and kitten states. When $N=2$ and $M=1$, by calculating the above superposition coefficients and phase angles, we obtain the cat state as
\begin{eqnarray}
\vert \psi_{\text{cat}}(\tau=\pi)\rangle_{a}=\frac{1}{\sqrt{2}}\left(e^{-i\pi/4}\vert i\alpha\rangle_{a}+e^{i\pi/4}\vert-i\alpha\rangle_{a}\right).\label{catstageome}
\end{eqnarray}
When $N=3$ and $M=1$, we obtain a kitten state with three superposition components as
\begin{equation}
\vert\psi_{\text{kitten}}(\tau=2\pi/3)\rangle_{a}=\frac{1}{\sqrt{3}}\left(e^{i\pi/6}\vert \alpha\rangle_{a} -i\vert
\alpha e^{i\frac{2}{3}\pi }\rangle_{a} +e^{i\pi/6}\vert \alpha e^{i\frac{4}{3}\pi}\rangle_{a}\right).\label{kittenstageome}
\end{equation}

The generated cat state and kitten state can be characterized by plotting their Wigner functions. For mode $a$ with the density matrix $\rho_{a}$, its Wigner function is defined by
\begin{eqnarray}
W(\xi)=\frac{2}{\pi}\text{Tr}[D_{a}^{\dagger}(\xi)\rho_{a}D_{a}(\xi)(-1)^{a^{\dagger}a}],
\end{eqnarray}
where $\xi$ is a complex variable and $D_{a}(\xi)=\exp(\xi a^{\dag}-\xi^{\star} a)$ is the usual displacement operator for mode $a$. Corresponding to the above two states $\vert \psi_{\text{cat}}(\tau=\pi)\rangle_{a}$ and $\vert\psi_{\text{kitten}}(\tau=2\pi/3)\rangle_{a}$, their Wigner functions can be calculated as
\begin{eqnarray}
W_{\vert\psi_{\text{cat}}(\tau=\pi)\rangle_{a}}(\xi)=\frac{1}{\pi}\left[\exp(-2\vert\xi-i\alpha\vert^{2})+\exp(-2\vert\xi+i\alpha\vert^{2})
+2\exp(-2\vert\xi\vert^{2})\sin[4\text{Re}(\alpha\xi^{\ast})]\right],\label{Wigfuncatstaanaly}
\end{eqnarray}
and
\begin{eqnarray}
W_{\vert\psi_{\text{kitten}}(\tau=2\pi/3)\rangle_{a}}(\xi)&=&\frac{2}{3\pi}\left[\exp(-2\vert\xi-\alpha\vert^{2})+\exp(-2\vert\xi-\alpha e^{i\frac{2}{3}\pi}\vert^{2})+\exp(-2\vert\xi-\alpha e^{i\frac{4}{3}\pi}\vert^{2})\right.\nonumber\\
&&\left.+2\text{Re}\left[-i\exp[i\text{Im}(\xi\alpha^{\ast}-\xi\alpha^{\ast}e^{-i\frac{2}{3}\pi})] e^{-i\frac{\pi}{6}}\;_{a}\langle\xi-\alpha\vert-\xi+\alpha e^{i\frac{2}{3}\pi}\rangle_{a}\right.\right.\nonumber\\
&&\left.\left.+\exp[i\text{Im}(\xi\alpha^{\ast}-\xi\alpha^{\ast}e^{-i\frac{4}{3}\pi})]\;_{a}\langle\xi-\alpha\vert-\xi+\alpha e^{i\frac{4}{3}\pi}\rangle_{a}\right.\right.\nonumber\\
&&\left.\left.+i\exp[i\text{Im}(\xi\alpha^{\ast}e^{-i\frac{2}{3}\pi}-\xi\alpha^{\ast}e^{-i\frac{4}{3}\pi})]e^{i\frac{\pi}{6}}\;_{a}\langle\xi-\alpha e^{i\frac{2}{3}\pi}\vert-\xi
+\alpha e^{i\frac{4}{3}\pi}\rangle_{a}\right]\right],\label{Wigfunkittenstaanaly}
\end{eqnarray}
where the overlaps between the coherent states $|\alpha\rangle$ and $|\beta\rangle$ can be calculated using the formula
\begin{eqnarray}
\langle\alpha|\beta\rangle=\exp\left(-\frac{1}{2}|\alpha|^2+\beta\alpha^{*}-\frac{1}{2}|\beta|^2\right).
\end{eqnarray}
In Fig.~\ref{Wignerfuncclose}, we show the two Wigner functions when the coherent amplitude is taken as $\alpha=2$. Here we can see that there are two and three main peaks, which correspond to the two and three superposition components in the two states $\vert \psi_{\text{cat}}(\tau=\pi)\rangle_{a}$ and $\vert\psi_{\text{kitten}}(\tau=2\pi/3)\rangle_{a}$, defined in Eqs.~(\ref{catstageome}) and~(\ref{kittenstageome}), respectively. These main peaks can be well resolved because the magnitude of the coherent state components is sufficiently large. For genertation of quantum superposition of macroscopically distinct states, the coherent amplitudes of superposed coherent states usually should be larger than $2$. Moreover, in the areas between the main peaks, we can see some oscillation pattern caused by quantum interference effect. These oscillation patterns are the signature of quantum superposition.

\begin{figure}[tbp]
\center
\includegraphics[bb=58 1 735 328, width=1 \textwidth]{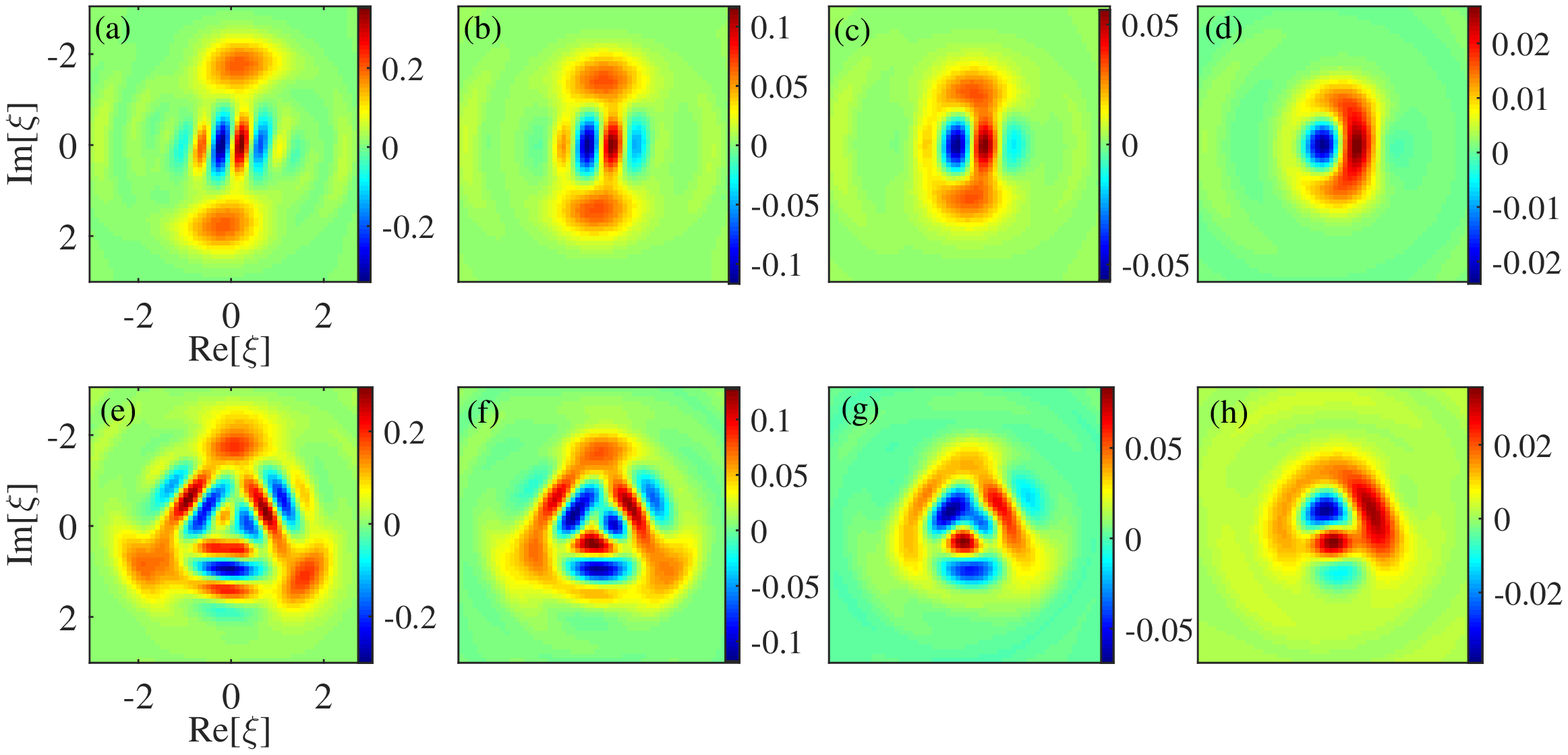}
\caption{(Color online) The Wigner functions correspond to the generated cat state and the kitten state in the open system case. Here panels (a)-(d) are the Wigner function for the Schr\"{o}dinger cat state with two superposition components, and panels (e)-(h) are the Wigner function for the Schr\"{o}dinger kitten state with three superposition components. The decay rates in these panels from left to right are given by $\gamma_{a}/\Delta_{b}=\gamma_{b}/\Delta_{b}=0.01$, $0.05$, $0.1$, and $0.2$. The coherent amplitude in the initial state $|\alpha\rangle$ is $\alpha=2$. Since the $\omega'_{a}$ term only introduces a rotation in the phase space, and hence the shape of the Wigner function is independent of $\omega'_{a}$. Other parameters are given by $\chi/\Delta_{b}=0.001$, $|\beta_{\textrm{ss}}|=1000$, $\bar{n}_{a}=\bar{n}_{b}=0$.}
\label{Wignerfuncdissip}
\end{figure}

\subsubsection{(b) The open-system case}

In this section, we study the influence of the dissipations on the generation of the cat state and the kitten state. To evaluate the influence of the dissipations, we need to divide the evolution as four steps, as explained in the above section. In principle, we can solve the four master equations corresponding to the evolution of the four steps. For the four steps, the Hamiltonians are different because the phase angles are different, and the final state of the previous step will be the initial state of the next step. However, in our simulations, we adopt the quantum trajectory method instead of the quantum mater equation. This is because the dimension of the Hilbert space is too large to be calculated with our computational resource. In our simulations, we choose $\alpha=2$ for the initial state of mode $a$. For the first three steps, the mode $b$ will be displaced by the generalized optomechanical coupling, and hence the dimension of the Hilbert space for mode $b$ should be taken as a large value (as estimated, $n_{b}$ should be larger than $120$ for high fidelity). For the master equation method, the number of the equations of motion is the square of the dimension of the total system, and hence the equations of motion become untreatable. Instead, by using the quantum trajectory method, the number of the equations of motion is largely decreased because it is equal to the dimension of the total Hilbert space. Below, we consider the zero-temperature case of the environments. In this case, the system can be described by a non-Hermite Hamiltonian by adding two imaginary terms phenomenologically as follows:
\begin{eqnarray}
H^{(l=1-4)}=\omega_{a}^{\prime}a^{\dagger}a-g_{0}a^{\dagger}a(b^{\dagger}e^{i\theta_{l}}+be^{-i\theta_{l}})-i\frac{\gamma_{a}}{2}a^{\dag}a-i\frac{\gamma_{b}}{2}b^{\dag}b,
\end{eqnarray}
where the phase angles are taken as $\theta_{1}=0$, $\theta_{2}=\pi/2$, $\theta_{3}=\pi$, and $\theta_{4}=3\pi/2$. Using these Hamiltonians, we can numerically solve the Schr\"{o}dinger equations step by step, then we can obtain the final state of the system. Based on the reduced density matrix of mode $a$, we can calculate the Wigner function of mode $a$.

To see the influence of the dissipation of the system on the cat state generation, in Fig.~\ref{Wignerfuncdissip} we plot the Wigner functions of the two- and three-component superposition states when the decay rates of the two modes take different values. Here, panels (a)-(d) correspond to the Wigner function of the cat state in the open-system case, while panels (e)-(h) correspond to the Wigner function of the kitten state in the open-system case. We can see that with the increase of the decay rate, the quantum interference evidence in the Wigner function disappears gradually. Therefore, the dissipation of the two modes will harm the quantum interference effect in the generated Schr\"{o}dinger cat and kitten states.

\section{IV. Discussions on the experimental implementation}

In this section, we present some discussions on the experimental implementation of this scheme with several possible candidate systems in quantum optics. The main result in this work  is the realization of a generalized ultrastrong optomechanical coupling in a cross-Kerr-type coupled two-mode system, in which one of the two modes is driven by a monochromatic field. As a result, the nominated physical systems should contain a cross-Kerr interaction, and one of the two modes should be driven by a monochromatic field. In addition, the parameter condition for this scheme is that the cross-Kerr parameter should be much smaller than the driving detuning $\Delta_{b}$ of mode $b$ such that the approximation used in discarding the cross-Kerr term in the transformed Hamiltonian is justified. For observing some quantum nonlinear effects in the ultrastrong coupling regime, we need a relatively large displacement amplitude $|\beta_{\text{ss}}|$ ($\sim10^{3}$ in our simulations) to enhance the generalized optomechanical coupling. Lastly, for resolving some sideband effects induced by mode $b$, the sideband-resolution condition $\Delta_{b}\gg\gamma_{a}$ should be satisfied. Therefore, the parameter conditions for implementation of this scheme are given by:
\begin{eqnarray}
\Delta_{b}\gg\chi,\hspace{1 cm}\Delta_{b}\gg\gamma_{a},\hspace{1 cm}g_{0}\equiv\chi\vert\beta_{\text{ss}}\vert\sim\Delta_{b}.
\end{eqnarray}
Note that the generalized optomechanical coupling strength $g_{0}$ should be much larger than the decay rate $\gamma_{a}$ of mode $a$, however, the cross-Kerr interaction strength $\chi$ could be either larger or smaller than the decay rate $\gamma_{a}$. This relaxed condition of the cross-Kerr interaction raises the possibility for implementation of this scheme with many quantum optical systems. In addition, since the driving detuning $\Delta_{b}$ (playing the role of the effective frequency of mode $b$ in the displacement representation) is a controllable parameter by choosing a proper driving frequency $\omega_{Lb}$, then we can design a proper $\Delta_{b}$ such that the relations $\Delta_{b}\gg\chi$ and $\Delta_{b}\gg\gamma_{a}$ are satisfied. We also choose a proper driving amplitude $\Omega_{b}$ such that the displacement amplitude $|\beta_{\text{ss}}|$ is large enough to confirm the relation $g_{0}\sim\Delta_{b}$.

In principle, our method is general and hence it can be implemented in various cross-Kerr-type coupled two-mode systems. In quantum optics, the cross-Kerr interaction between two bosonic modes is usually obtained by coupling these two modes to a common intermediate, which could be a three-level atom~\cite{Liu2017}, an ``$N$"-type atom~\cite{Schmidt1996,Kang2003,Sinclair2007,Sinclair2008}, and an ``$M$"-type atom~\cite{Matsko2003}, where the atom could be either a natural or an artificial atom, which corresponds to cavity-QED~\cite{Kimble1998} or circuit-QED~\cite{Hu2011,Nigg2012,Bourassa2012,Holland2015,Majer2007,Hoi2013}. We also discuss the implementation of this cross-Kerr interaction in a quadratic optomechanical system with a ``membrane-in-the-middle" configuration~\cite{Thompson2008,Sankey2010,Karuza2013} and in a coupled cavity-ion system~\cite{Semiao2005,Maurer2004}. Below, we will present some analyses on the current experimental conditions in these systems. For the purpose of comparison, we list in Table~\ref{tablesm1} some relating parameters reported in either theoretical proposals or realistic experiments concerning the two-mode cross-Kerr-type-coupled systems. Note that great advances have been made in circuit-QED~\cite{Hu2011,Nigg2012,Bourassa2012,Holland2015,Majer2007,Hoi2013}. From Table~\ref{tablesm1} we can see that depending on the task, the coupled two-mode systems can be designed to have a wide range of parameters.
\begin{table}[]
\centering
\caption{Parameters of the cross-Kerr-type coupled systems reported in literature: the resonance frequencies $\omega_{a}$ and $\omega_{b}$ of modes $a$ and $b$, the cross-Kerr interaction strength $\chi$, the decay rates $\gamma_{a}$ and $\gamma_{b}$ for modes $a$ and $b$, the thermal excitation occupations $\bar{n}_{a}$ and $\bar{n}_{b}$ in the baths of modes $a$ and $b$. Here the subscripts $``E"$ and $``T"$ of the reference number denote the reference as an experimental work and a theoretical work, respectively.}
\label{tablesm1}
\begin{tabular}{|c|c|c|c|c|c|c|c|c|}
\hline
Ref. & description &  $\frac{\omega_{a}}{2\pi}$ (GHz) & $\frac{\omega_{b}}{2\pi}$ (GHz) &  $\frac{\chi}{2\pi}$ (kHz)  & $\frac{\gamma_{a}}{2\pi}$ (kHz)& $\frac{\gamma_{b}}{2\pi}$ (kHz)  & $\bar{n}_{a}$ & $\bar{n}_{b}$ \\
\hline
\cite{Holland2015}$_{E}$ &  circuit-QED & $8.493$ & $9.32$ & $2.59\times10^{3}$ & $1.25$   & $5.25$  & $\sim0$ & $\sim0$ \\
\hline
\cite{Hu2011}$_{T}$, \cite{Majer2007} &  circuit-QED  & $\sim5$  & $\sim5$ & $2.5\times10^{3}$ & $\sim30\times10^{3}$ & $\sim30\times10^{3}$ &  $\sim0$  & $\sim 0$ \\
\hline
\cite{Sankey2010}$_{E}$ &quadratic optomech. & $282\times10^{3}$   &  $0.134\times10^{-3}$  &  $0.395\times10^{-4}$  &  $5.94\times10^{2}$  &  $ 0.122\times10^{-3}$ & $ \sim0  $&  $46648$  \\
\hline
\cite{Semiao2005}$_{T}$,\cite{Maurer2004} & cavity-ion system  &  $351\times10^{3}$  &  $\gg15.1\times10^{-3}$  &  $0.796$  &  $41.7$   &    &  $\sim0$ &\\
\hline
\end{tabular}
\end{table}

Some proposals on the cross-Kerr interaction are based on the system consisting of two continuous-wave optical fields coupled to an ensemble of atoms with an ``N"-type configuration~\cite{Schmidt1996,Kang2003,Sinclair2007,Sinclair2008}. This method also works for the cavity field case and hence a cross-Kerr interaction between two cavity fields can be obtained. For optical cavity modes, the resonance frequencies ($\sim10^{14}$ Hz) are much larger than the decay rates ($\sim10^{7}$ Hz), and the thermal occupation numbers are negligible. In typical optical cavity-QED systems, the atom-field coupling strength is of the order of $10^{8}$ Hz~\cite{Kimble1998}. Since the cross-Kerr interaction is of the order of the atom-field coupling strength times the cube of a small ratio (the value of the ratio is much smaller than one, for example $10^{-1}$), then the effective cross-Kerr parameter could be of the order of $10^{4}$-$10^{5}$ Hz. For microwave field modes, the resonance frequencies might be of the order of $5$ - $10$ GHz and the decay rates are $10^{4}$ - $10^{5}$ Hz. The thermal occupation numbers are negligible at the temperature around $20$ mK. The cross-Kerr parameters are of the order of $10^{6}$ Hz.

(i) If we consider the case where the two bosonic modes in our model are electromagnetic fields, then the sideband-resolution condition is satisfied. In addition, we choose proper driving frequency $\omega_{Lb}$ such that $\Delta_{b}=\omega_{b}-\omega_{Lb}\gg\gamma_{a}$ and $\Delta_{b}\gg\chi$, then the sideband-resolution condition is satisfied and the approximation is justified. We also choose a proper driving amplitude $\Omega_{b}$ such that $g_{0}=\chi|\beta_{\text{ss}}|\sim\Delta_{b}$.

(ii) If the coupled two-mode system is composed by an electromagnetic field and a mechanical mode, then the resonance frequency of mode $b$ is of the order of $10^{7}$ - $10^{8}$ Hz, and the decay rate of mechanical mode is of the order of $10^{2}$ - $10^{3}$ Hz. In this case, the system can still work in the resolved-sideband regime. The coupling strength $g_{0}$ can also be enhanced to be larger than the decay rate $\gamma_{a}$ and the effective frequency $\Delta_{b}$. However, the thermal occupation number $\bar{n}_{b}$ in the mechanical resonator case will be a finite number (dozens of thermal phonons).

(iii) For the quadratic optomechanical systems, the quadratic optomechanical coupling can be approximated as a cross-Kerr interaction. However, the magnitude of the cross-Kerr interaction is small. For obtaining a $g_{0}\sim\Delta_{b}\gg\{\chi,\gamma_{a}\}$, the displacement $|\beta_{\text{ss}}|$ needs to be a very large number.

(iv) For a coupled cavity-ion system, the sideband-resolution condition is satisfied, and the cross-Kerr parameter $\chi$ takes a moderate value. It can be enhanced to be larger than $\Delta_{b}$ and $\gamma_{a}$ by designing a displacement $|\beta_{\text{ss}}|$. For this system, a key point is to suppress the thermal noise as much as possible.

\begin{table}[]
\centering
\caption{The parameters used in our simulations: the resonance frequency $\omega'_{a}=\omega_{a}+\chi|\beta_{\text{ss}}|^2$ of mode $a$, the driving detuning (the effective frequency in the transformed representation) $\Delta_{b}=\omega_{b}-\omega_{Lb}$ of mode $b$, the cross-Kerr interaction strength $\chi$, the displacement amplitude $|\beta_{\text{ss}}|$, the single-photon optomechanical-coupling strength $g_{0}=\chi|\beta_{\text{ss}}|$, the decay rates $\gamma_{a}$ and $\gamma_{b}$ of modes $a$ and $b$, the thermal occupation numbers $\bar{n}_{a}$ and $\bar{n}_{b}$ in the baths of modes $a$ and $b$, and the single-photon strong-coupling condition $g_{0}/\gamma_{a}$.}
\label{table2sm}
\begin{tabular}{|c|c|c|c|}
\hline
Notation & Remarks & Scaled parameters  &Parameters  \\
\hline
$\omega'_{a}$ & arbitrary &   &  \\
\hline
$\Delta_{b}$ & as the frequency scale  & $1$ & $2\pi\times 1$ MHz \\
\hline
$\chi$   & $\chi/\Delta_{b}\ll1$ for approximation &  $\chi/\Delta_{b}=0.001$ - $0.01$ & $2\pi\times$ ($1$ - $10$) kHz \\
\hline
$|\beta_{\text{ss}}|$ & $|\beta_{\text{ss}}|\gg1$ for coupling enhancement  &  & $1000$ - $2000$ or $100$ - $200$ \\
\hline
$g_{0}=\chi|\beta_{\text{ss}}|$ &  enhanced optomechanical-coupling strength  &  $g_{0}/\Delta_{b}\sim1$ - $2$ & $2\pi\times$ ($1$ - $2$) MHz  \\
\hline
$\gamma_{a}$  & decay rate of mode $a$ & $\gamma_{a}/\Delta_{b}=0.01$ - $0.1$ & $2\pi\times$ ($10$ - $100$) kHz \\
\hline
$\gamma_{b}$  & decay rate of mode $b$  & $\gamma_{b}/\Delta_{b}=0.01$ - $0.1$   &  $2\pi\times$ ($10$ - $100$) kHz \\
\hline
$\bar{n}_{a}$  & negligible for optical and microwave fields  &   &  $0$  \\
\hline
$\bar{n}_{b}$ & a finite number for a mechanical resonator  &    & $0$ - $10$   \\
\hline
$g_{0}/\gamma_{a}$ & single-photon strong-coupling condition  &  $10$ - $200$  &    \\
\hline
\end{tabular}
\end{table}

Based on the above discussions, in Table~\ref{table2sm} we suggest some parameters for simulation of this scheme. In our model, the involved parameters include: the resonance frequencies $\omega_{a}$ and $\omega_{b}$, the cross-Kerr interaction strength $\chi$, the decay rates of the two bosonic modes $\gamma_{a}$ and $\gamma_{b}$, dimensionless displacement amplitude $|\beta_{\text{ss}}|$, the driving amplitude $\Omega_{b}$ and frequency $\omega_{Lb}$ of mode $b$. Below we analyze the feasibility of our scheme based on the above listed parameters.
Mode $a$ could be either optical or microwave mode. In the cat state generation tasks, the free Hamiltonian of mode $a$ will not affect the dynamics of the system, because it commutates the other terms in the Hamiltonian. For the photon blockade task, the single photon resonance is taken and then useful parameter is the driving detuning $\Delta_{a}=\omega_{a}-\omega_{La}$. For optical mode $a$, its frequency is of the order of hundreds of terahertz, the decay rate might be $\gamma_{a}\sim2\pi\times10$ - $100$ MHz. For microwave mode $a$, its frequency might be $\omega_{a}\sim2\pi\times5$ - $10$ GHz, the decay rate might be $\gamma_{a}\sim2\pi\times100$ kHz. The cross-Kerr interaction between the two modes is of the order of $1$ - $10$ kHz~\cite{Holland2015}. By choosing proper driving amplitude and frequency, the system can work in the ultrastrong-coupling regime.

\end{document}